\documentclass[11pt]{article}
\usepackage{epsf,amsmath,amssymb,graphicx,scalefnt,rotating,enumitem,fancyhdr}
\usepackage{menukeys}
\usepackage{array}
\usepackage{acronym}

\usepackage{filemod}
\usepackage{hyperref}
\usepackage[affil-it]{authblk}
\usepackage{slashed}
\usepackage{dsfont}

\usepackage{listings}

\usepackage[T1]{fontenc}
\usepackage[utf8]{inputenc}
\usepackage{tikz}
\usepackage[final]{showlabels}

\usetikzlibrary{backgrounds}
\usetikzlibrary{calc}

\definecolor{codegreen}{rgb}{0,0.6,0}
\definecolor{codegray}{rgb}{0.5,0.5,0.5}
\definecolor{codepurple}{rgb}{0.58,0,0.82}
\definecolor{backcolour}{rgb}{0.95,0.95,0.92}

\lstdefinestyle{mystyle}{
    backgroundcolor=\color{backcolour},
    commentstyle=\color{codegreen},
    keywordstyle=\color{magenta},
    numberstyle=\tiny\color{codegray},
    stringstyle=\color{codepurple},
    basicstyle=\ttfamily\footnotesize,
    breakatwhitespace=false,
    breaklines=true,
    captionpos=b,
    keepspaces=true,
    numbers=left,
    numbersep=5pt,
    showspaces=false,
    showstringspaces=false,
    showtabs=false,
    tabsize=2
}

\newcounter{notecount}
\newcommand{\levelfile}{\textit{level file}}
\newcommand{\modelfile}{\textit{model file}}
\newcommand{\infin}{\texttt{InFin}}
\newcommand{\editframe}{\textit{EditFrame}}
\newcommand{\keystroke}[1]{\keys{#1}}
\newcommand{\editframekey}{\keystroke{e}}
\newcommand{\helperkey}{\keystroke{h}}
\newcommand{\invertkey}{\keystroke{i}}
\newcommand{\activekey}{\keystroke{a}}
\newcommand{\gridkey}{\keystroke{g}}

\newcommand{\feyngame}{\texttt{FeynGame}}
\newcommand{\note}[2]{ {\stepcounter{notecount}\sf%
    \footnotesize\color{red}[\arabic{notecount}$|${\color{blue}#1}$|$#2]}}
\renewcommand{\note}[2]{}
\newcommand{\citere}[1]{Ref.\,\cite{#1}}

\newcommand{\abbrev}[1]{{\scalefont{.9}#1}}
\newcommand{\ep}{\epsilon}

\newcommand{\fig}[1]{Fig.\,\ref{#1}}

\newcommand{\rhtab}[1]{Table\,\ref{#1}}

\newcommand{\sct}[1]{Sect.\,\ref{#1}}

\newcommand{\dd}{\mathrm{d}}

\newcommand{\deriv}[3]{\frac{\partial\ifthenelse{\equal{#1}{}}{}{^{#1}}%
    #2}{\partial #3\ifthenelse{\equal{#1}{}}{}{^{#1}}}}
\newcommand{\dderiv}[3]{\frac{\dd\ifthenelse{\equal{#1}{}}{}{^{#1}}%
    #2}{\dd #3\ifthenelse{\equal{#1}{}}{}{^{#1}}}}

\newcommand{\myacrodef}[3]{\acrodef{#2}{#3}\newcommand{#1}{\ac{#2}}}
\myacrodef{\qcd}{QCD}{Quantum Chromo Dynamics}
\acused{QCD} 
\myacrodef{\qed}{QED}{Quantum Electrodynamics}
\acused{QED} 
\myacrodef{\pdf}{PDF}{}
\acused{PDF}
\myacrodef{\lhc}{LHC}{Large Hadron Collider}
\myacrodef{\gui}{GUI}{graphical user interface}
\myacrodef{\uv}{UV}{ultra-violet}
\myacrodef{\lo}{LO}{leading order}
\myacrodef{\nlo}{NLO}{next-to-leading order}
\myacrodef{\nnlo}{NNLO}{next-to-next-to-leading order}
\myacrodef{\llog}{LL}{leading logarithmic}
\myacrodef{\nll}{NLL}{next-to-leading logarithmic}
\myacrodef{\nnll}{NNLL}{next-to-next-to-leading logarithmic}
\myacrodef{\sm}{SM}{Standard Model}
\myacrodef{\bsm}{BSM}{beyond-the-\ac{SM}}
\myacrodef{\mssm}{MSSM}{Minimal Supersymmetric \ac{SM}}
\myacrodef{\susy}{SUSY}{Supersymmetry}
\myacrodef{\dreg}{DREG}{Dimensional Regularization}
\myacrodef{\dred}{DRED}{Dimensional Reduction}
\myacrodef{\emt}{EMT}{energy-momentum tensor}

\newcommand{\myheaderline}{TTK-20-04 --- February 2020}
\fancypagestyle{firstpage}
{
  
  \fancyhead[R]{\myheaderline}
}
\title{
  \includegraphics[viewport = 18 41 541
    824,height=.2\textheight]{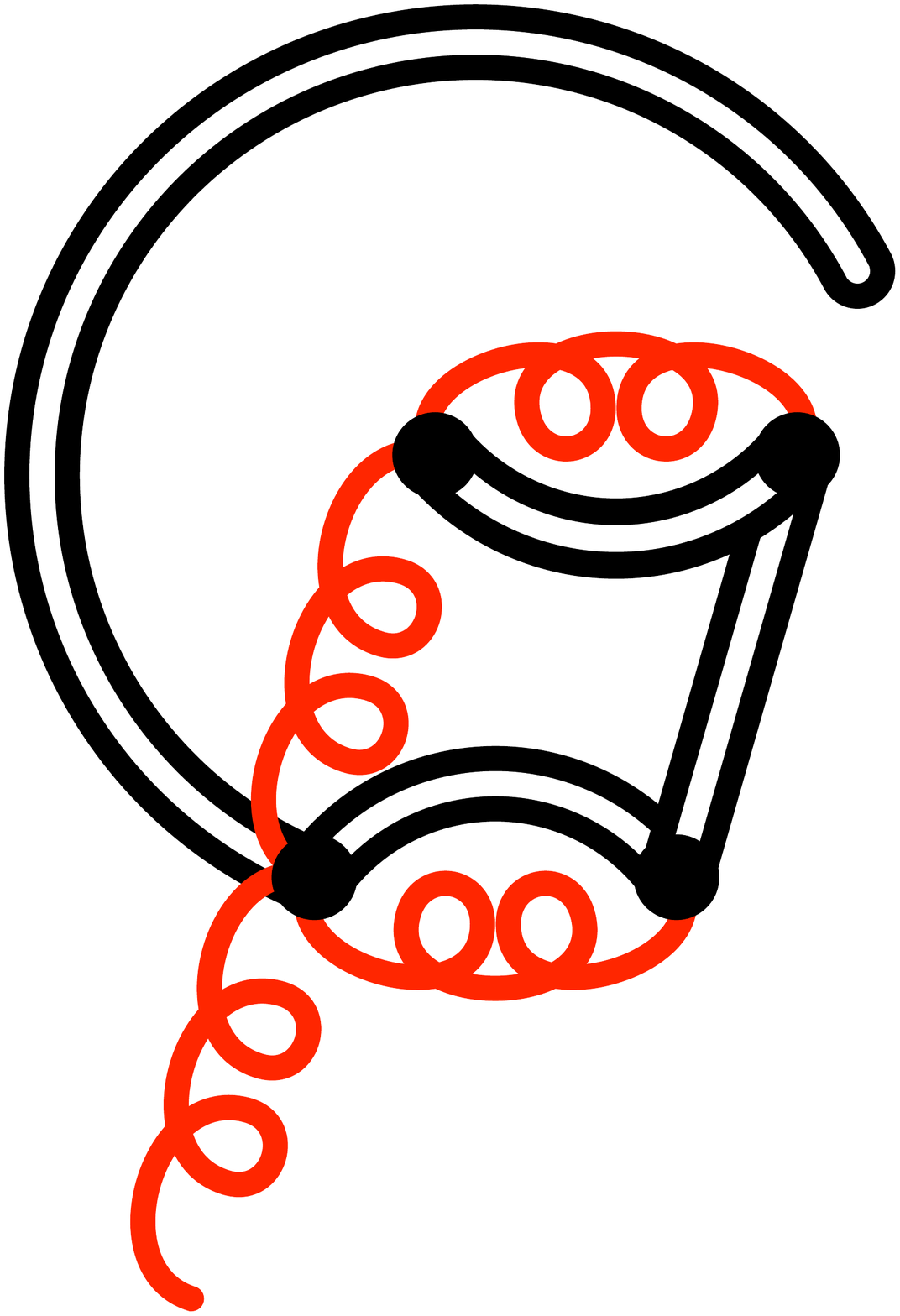}\\
  FeynGame}
\author{R.V. Harlander, S.Y. Klein, M. Lipp}
\affil{Institute for Theoretical Particle Physics and Cosmology,\\ RWTH
  Aachen University, 52074 Aachen, Germany}
\date{}
\begin{document}
\maketitle
\begin{abstract}
  A \texttt{java}-based graphical tool for drawing Feynman diagrams is
  presented. It differs from similar existing tools in various
  respects. For example, it is based on models, consisting of particles
  (lines) and (optionally) vertices, each of which can be given their
  individual properties (line style, color, arrows, label, etc.). The
  diagrams can be exported in any standard image format, or as
  \pdf. Aside from its plain graphical aspect, the goal of \feyngame\ is
  also educative, as it can check a Feynman diagrams validity. This
  provides the basis to play games with diagrams, for example. Here we
  describe on such game where a given set of initial and final states
  must be connected through a Feynman diagram within a given interaction
  model.
\end{abstract}
\thispagestyle{firstpage}



\newpage

\subsection*{PROGRAM SUMMARY}

\textit{Program title:} \feyngame\\
\textit{Distribution format:} \href{http://www.robert-harlander.de/software/feyngame}{gzipped tar archive},
\href{https://gitlab.com/feyngame/FeynGame}{GitLab repository}\\
\textit{Authors:} R.V. Harlander, S.Y. Klein, M. Lipp\\
\textit{Licensing provisions:} GNU General Public License 3 (GPL)\\
\textit{Programming language:} \texttt{Java}\\
\textit{Operating systems:} Linux, Windows, MacOS\\
\textit{Keywords:} Feynman diagrams, Java, GUI\\
\textit{Nature of problem:} Efficient drawing of Feynman diagrams for
presentations and publications; playful elementary introduction to the
concept of Feynman diagrams \\
\textit{Method of solution:} Graphical interface which incorporates the
Feynman rules for various models of particle interactions. \\
\textit{Restrictions:} Only the topological information of the Feynman rules is
incorporated.\\
\textit{Running time:} The running time depends on the diagram
to be drawn and the skill and practice of the user.

\newpage

\section{Introduction}\label{sec:intro}

Feynman diagrams are a central tool for particle
physics\,\cite{Feynman:1949hz}. Not only are they indispensable for
perturbative calculations of observables such as cross sections or decay
rates. They are also extremely useful in every-day scientific
communication, from the professional research to the educational and
even the popular science level (see, e.g.,
\citere{Kaiser:2005aa,Wuthrich:2010aa}). In this way, they provide an
excellent bridge to convey scientific knowledge to a broad audience.

A perturbative calculation for a given process within a specific model
(e.g.\ the \sm) typically starts with the generation of the relevant
Feynman diagrams. This is an algorithmic process which can be
implemented into a computer program. The most prominent examples for
such implementations are
\texttt{FeynArts}\,\cite{Hahn:2000kx,Kublbeck:1992mt} and
\texttt{qgraf}\,\cite{Nogueira:2006pq}. The diagrams are then usually
passed on to other programs which translate them into mathematical
expressions according to the so-called Feynman rules, and yet again to
other tools which actually evaluate these expressions. A paradigm
example where all these steps are incorporated up to \nlo{} in
perturbation theory in a single framework is
\texttt{MadGraph}\,\cite{Alwall:2011uj}.

In all of this process chain, the actual visualization of the diagrams
is not required, since in the ideal case no human intervention is
necessary. Only the computer ``sees'' the diagrams, albeit in some
mathematical encoding such as incidence matrices. Nevertheless,
communication about particle physics still relies heavily on Feynman
diagrams, for example in order to refer to a particular process or
specific contributions to it (see, e.g., \citere{Stoeltzner:2018aa}). On
the other hand, their one-to-one correspondence to Feynman
\textit{integrals} also increases the readability of relations among
these integrals. For example, a famous identity among massless scalar
integrals takes the form \cite{Chetyrkin:1981qh}
\begin{equation}
  \begin{split}
   \ep\,\raisebox{-2.2em}{\includegraphics[viewport = 18 517 577 824,clip,
       width=.22\textwidth]{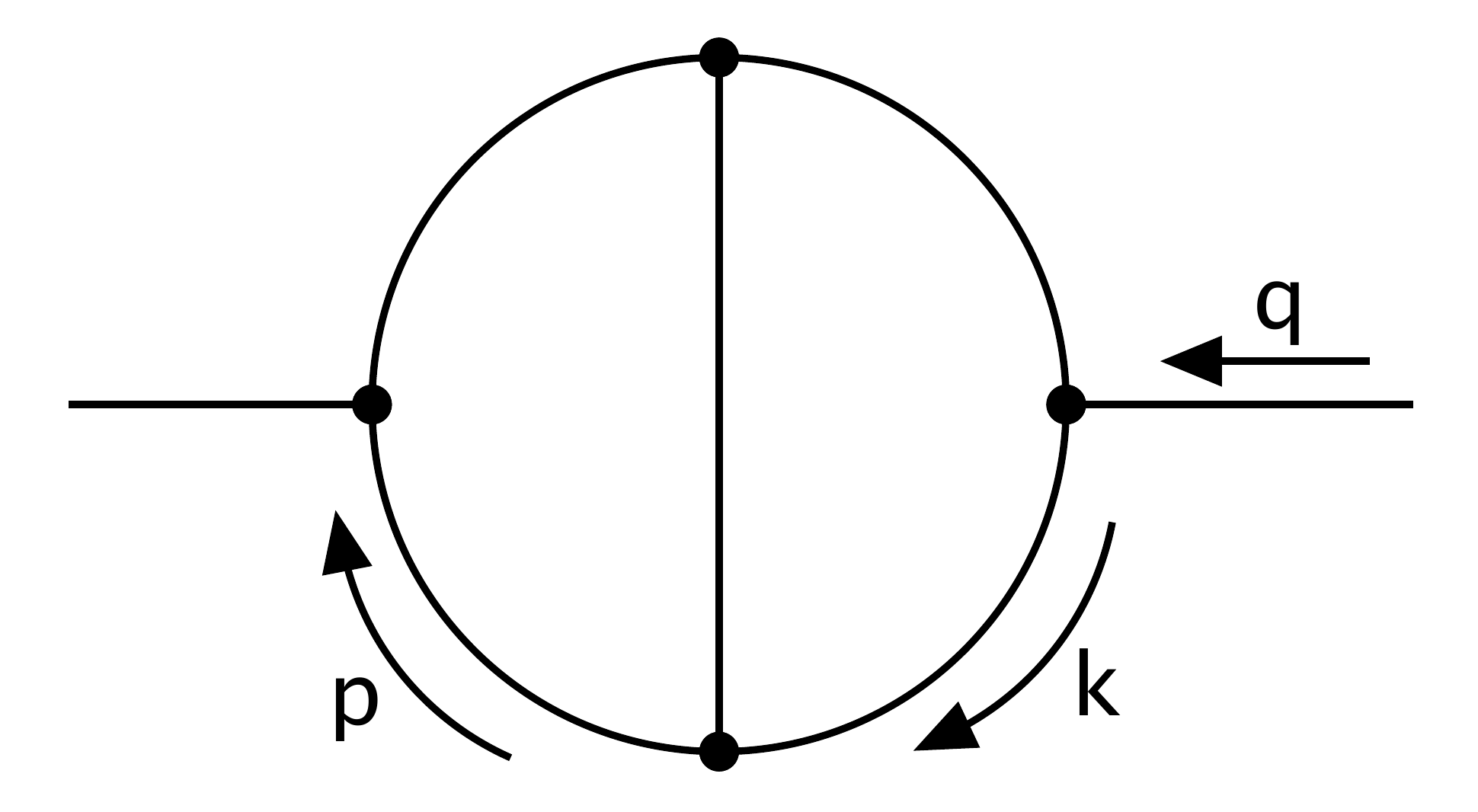}}\
   = \raisebox{-2.2em}{\includegraphics[viewport = 18 517 577 824,clip,
         width=.22\textwidth]{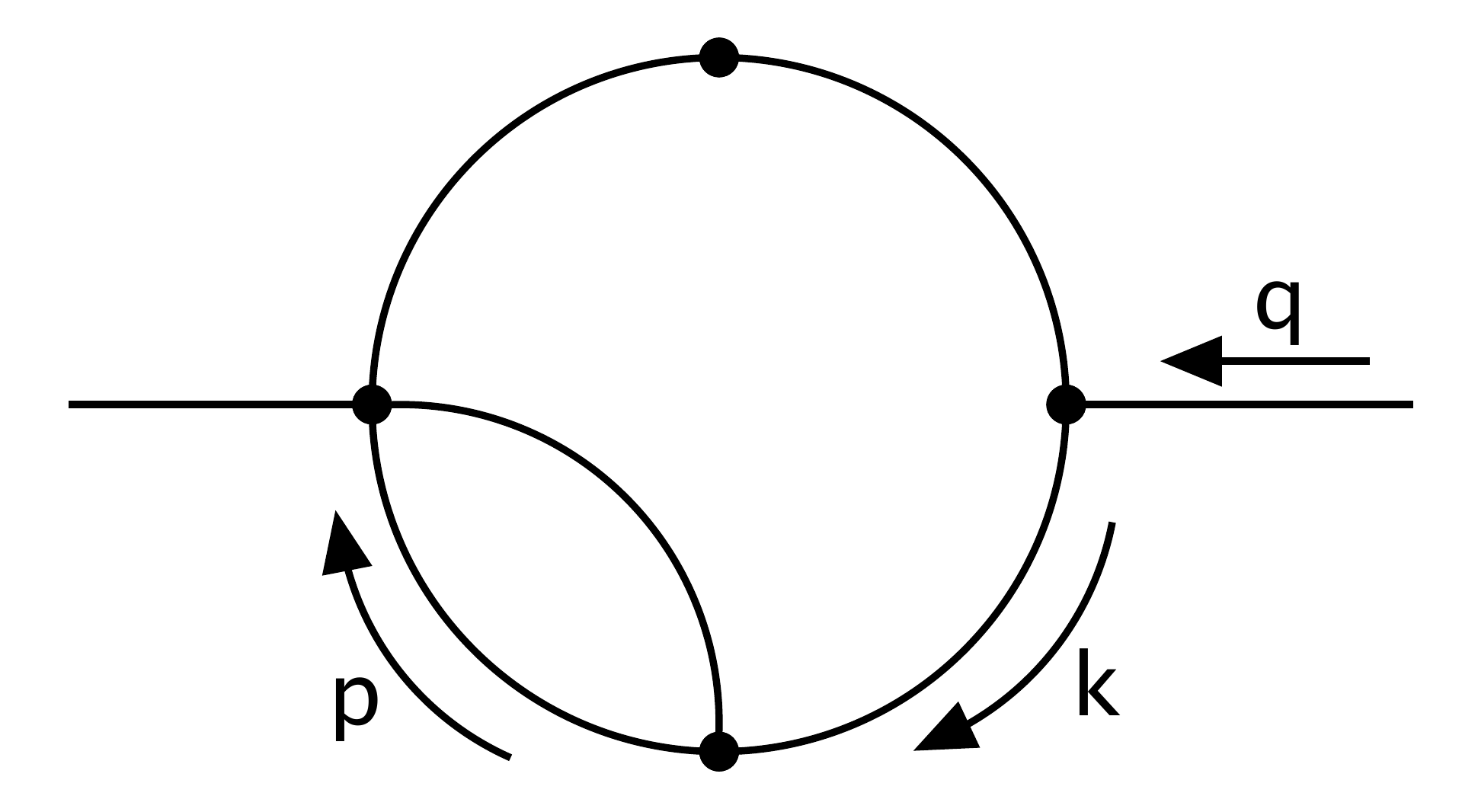}}\ -
     \raisebox{-3.25em}{\includegraphics[viewport = 18 517 577 824,clip,
         width=.22\textwidth]{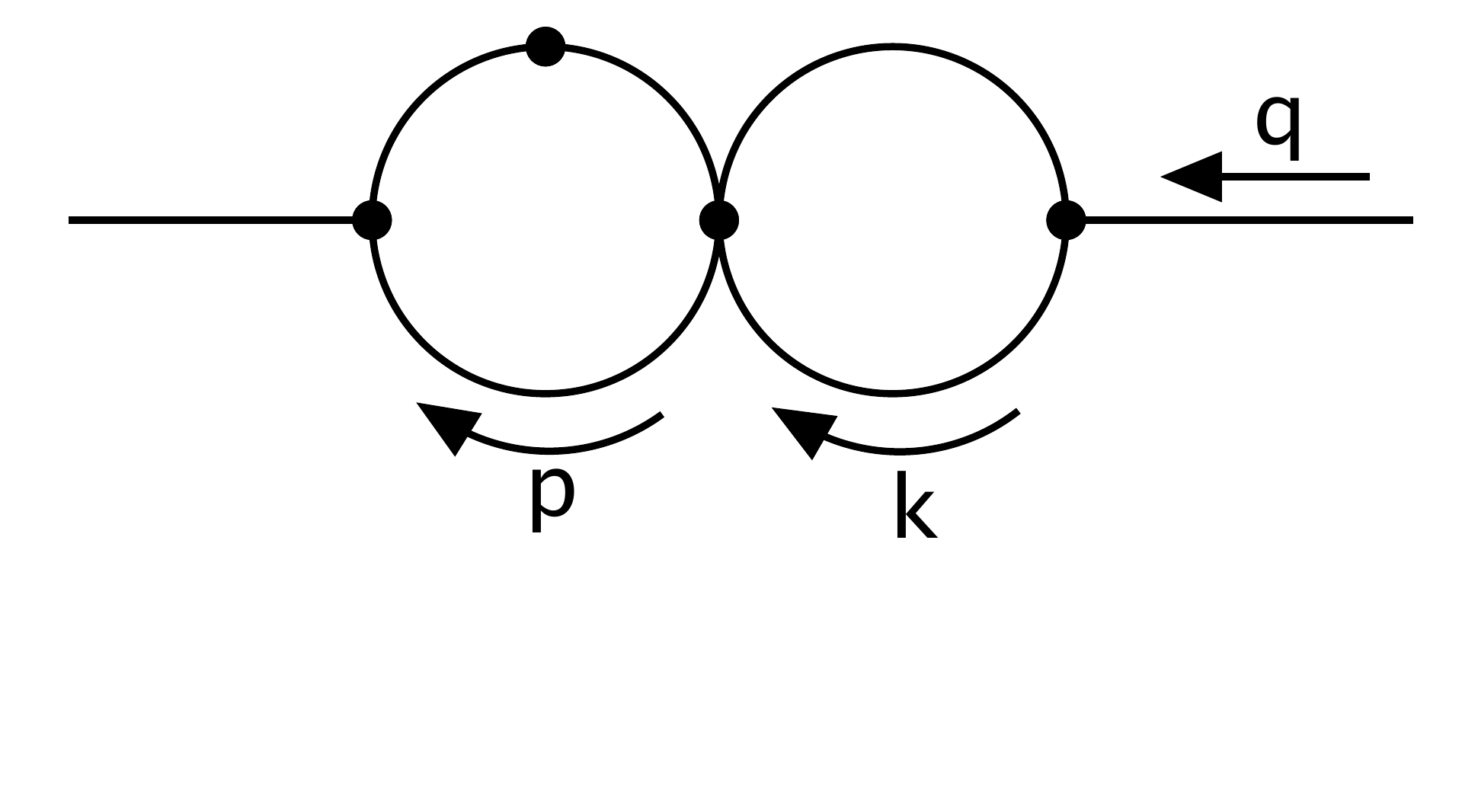}}\,,
  \end{split}
\end{equation}
which the expert reader immediately translates into Feynman
integrals:
\begin{equation}
  \begin{split}
    \ep\int\frac{ \dd^Dp\,\dd^Dk}{p^2k^2(k-q)^2(p-q)^2(k-p)^2} =\\
    =\int\frac{ \dd^Dp\,\dd^Dk}{p^2k^2(k-q)^4(k-p)^2}
    -\int\frac{ \dd^Dp\,\dd^Dk}{p^2k^2(k-q)^2(p-q)^4}
    \label{eq:}
  \end{split}
\end{equation}
which holds for space-time dimensions $D=4-2\ep$, with $\ep\neq 0$.

For these reasons, computer tools for the actual \textit{visualization}
of Feynman diagrams are important.  It is probably fair to say that the
discussion about the most suitable tool for this purpose is one of
physicists' favorite (non-scientific) controversial subjects. It seems
that the optimal compromise between handiness, flexibility, and
esthetics has not been achieved yet by any of the existing computer
tools that have been designed for this purpose.

The latter naturally fall into two categories: text-based and graphical
tools. The former usually provide a set of routines and statements
(``packages'') within a framework like \LaTeX\ or \texttt{C++} which
encode graphical objects. The user needs to write some code which, upon
compilation, produces a visual image of the diagram. Graphical tools, on
the other hand, provide a \gui\ which allows the user to directly draw
the diagrams onto a ``canvas'' with the ``mouse''.\footnote{We use the
  expressions ``mouse'', ``mouse wheel'', ``click'' etc.\ in a generic
  way, referring to all kinds of ``mouse-like'' input devices such as
  track pads or tablet pens, and the corresponding control actions.}

The earliest examples for text-based tools are probably
\texttt{FeynDiagram}\,\cite{FeynDiagram}, which is a set of \texttt{C++}
objects for producing Feynman diagrams in \texttt{PostScript} format, or
\texttt{axodraw}\,\cite{Vermaseren:1994je} and
\texttt{FeynMF/FeynMP}\,\cite{Ohl:1995kr,FeynMP}, both of which allow to
include the code directly into the source of a \LaTeX\ document. The
diagram is then generated by compiling the \LaTeX\ code, possibly with
accompanying runs of additional programs such as \texttt{metafont},
\texttt{metapost}, or \texttt{ps2pdf}. More recent tools are
\texttt{axodraw2}\,\cite{Collins:2016aya} and
\texttt{PyFeyn}\,\cite{PyFeyn} while currently the most popular
text-based system seems to be
\texttt{TikZ-Feynman}\,\cite{Ellis:2016jkw}, which uses
\texttt{lualatex}\,\cite{lualatex} to automatically position the
vertices properly.

The most widely used tool which includes a \gui\ is probably
\texttt{JaxoDraw}\,\cite{Binosi:2008ig}, which is a \texttt{java}
interface for \texttt{axodraw}.  Recently, a few other graphical tools
have appeared, most notably \texttt{feynman}\cite{Feynman} and
\texttt{Feynman diagram maker}\,\cite{FeynMaker}, both of which can be
run online from within a browser.

All of the tools above, whether text- or \gui-based, are designed to
draw individual Feynman diagrams ``line-by-line''. There are only very
few tools that automatically visualize the output of Feynman diagram
generators. The prototype of such tools has been
\texttt{FeynArts}\,\cite{Kublbeck:1992mt,Hahn:2000kx},
which can even be combined with the \texttt{FeynEdit}
\gui\,\cite{Hahn:2007ue} in order to feyn-, sorry: fine-tune the output
of \texttt{FeynArts} (it can also be used stand-alone).

The only tool to produce animated Feynman diagrams that we are aware of
is \texttt{aximate}\,\cite{aximate}.

Rather than doing a detailed comparison of \feyngame\ with existing
tools, let us simply state what we believe are its most important, and
in particular its unique features. The purpose of \feyngame\ is
two-fold: on the one hand, it should allow to draw Feynman diagrams of
high quality in a very fast manner. It is thus well-suited for quickly
supplying \texttt{Keynote} or \texttt{PowerPoint} presentations with
Feynman diagrams through a simple cut-and-paste operation. By exporting
the diagrams as \pdf, they can of course be included in type-set
documents such as books or scientific papers as well.

The second purpose of \feyngame\ is educational: since its usage is
based on particle models (\qed, \sm, etc.), it can check whether a
particular Feynman diagram is consistent with the underlying model. This
provides the basis for playing games with Feynman diagrams. The current
version of \feyngame\ contains the game \texttt{InFin}, where the player
is asked to turn a random pair of initial and final states into a valid
connected Feynman diagram. We believe that this may help to bring the
concept of Feynman diagrams closer to high-school students or
undergraduates. Forthcoming versions of \feyngame\ will include further
games.

\feyngame\ is used through a \gui\ written in \texttt{java}. It is
therefore highly portable and should run on any platform which provides
\texttt{java} (version 8 or later).  Even though \feyngame\ should work
out-of-the-box, it is designed to be personalized. This means that the
user can provide her own \modelfile\ which defines all the
desired (or required) line and vertex styles. This avoids the need to
modify every individual line or vertex in case its appearance is
required to deviate from the overall default. For example,
\feyngame\ provides a \modelfile\ for the \sm\ which contains a unique
line for each particle of that model. To draw a particular Feynman
diagram, one simply picks the individual lines which are displayed in
the main window of \feyngame.  The user can conveniently change the
overall appearance of these lines, either through the \gui\ itself, or
by editing the \modelfile.


%
\begin{figure}
  \begin{center}
    \begin{tabular}{ccc}
      \includegraphics[viewport=60 80 650 580, width=.3\textwidth]{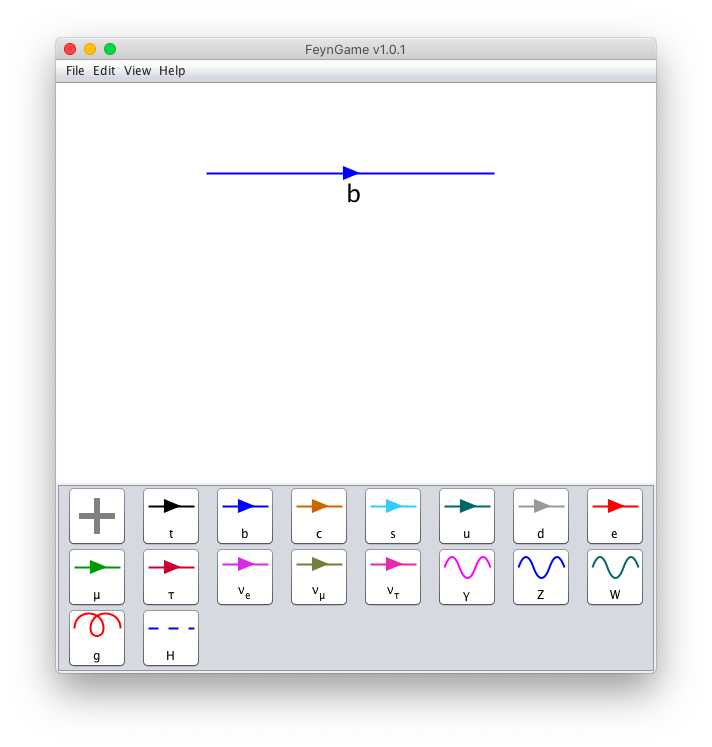} &
            \includegraphics[viewport=60 80 650 580, width=.3\textwidth]{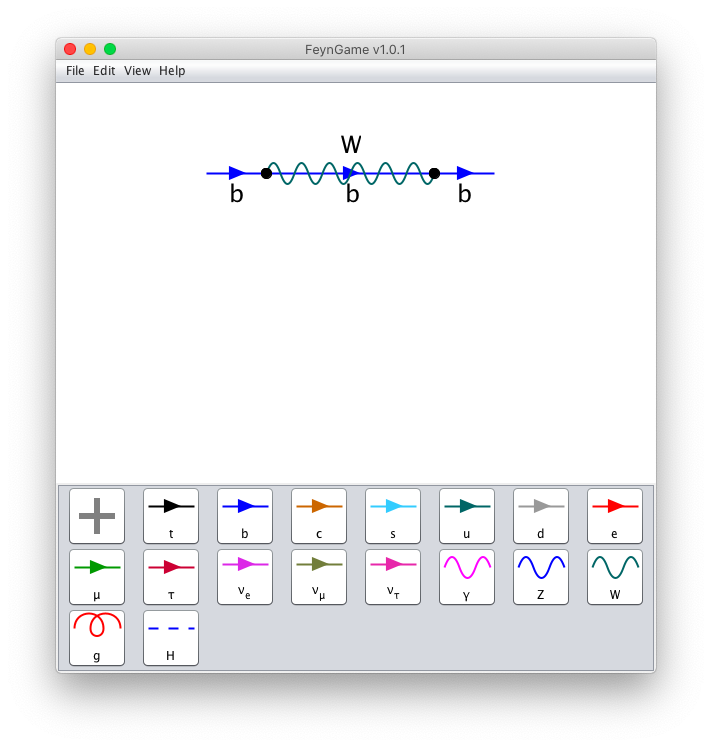} &
            \includegraphics[viewport=60 80 650 580,
              width=.3\textwidth]{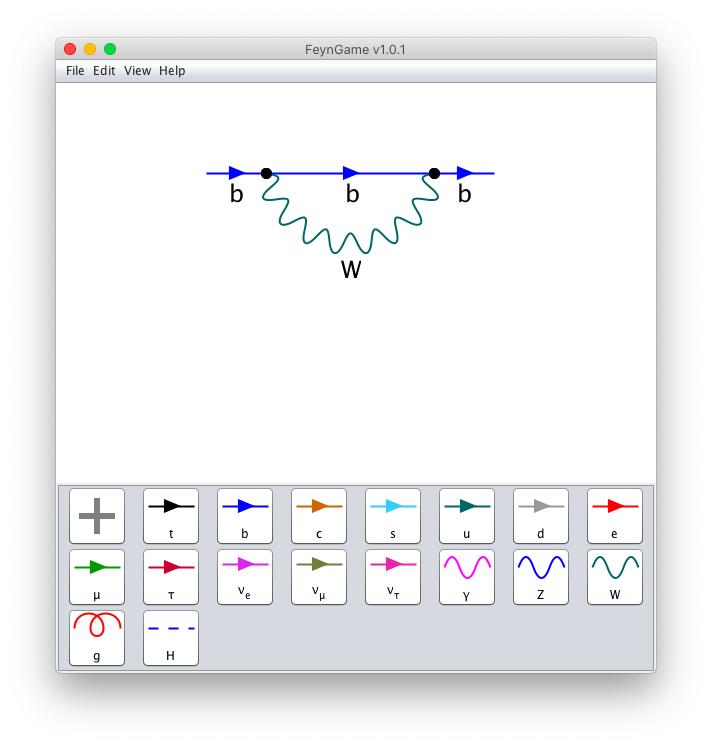}
                        \\
            (i) & (ii) & (iii)\\[2em]
            \includegraphics[viewport=60 80 650 580,
              width=.3\textwidth]{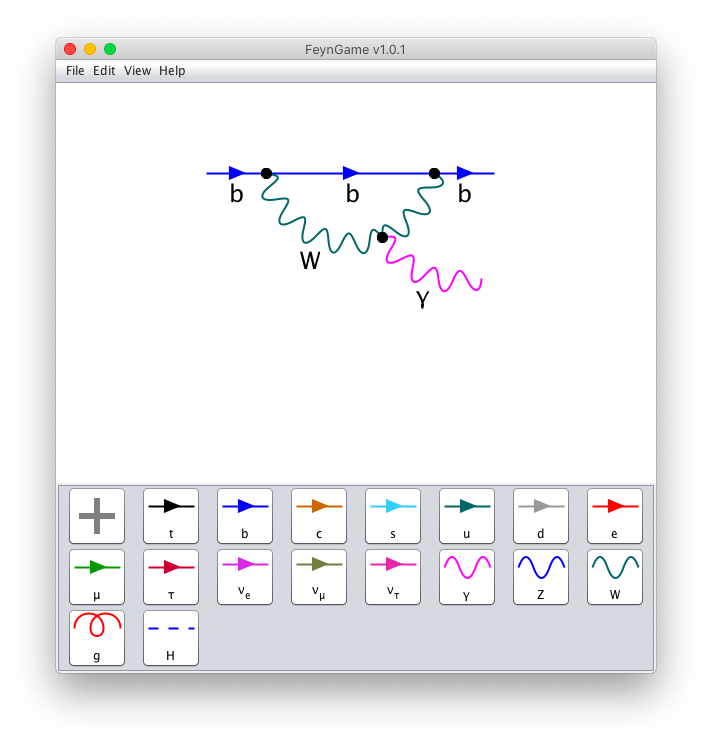} &
            \includegraphics[viewport=60 80 650 580, width=.3\textwidth]{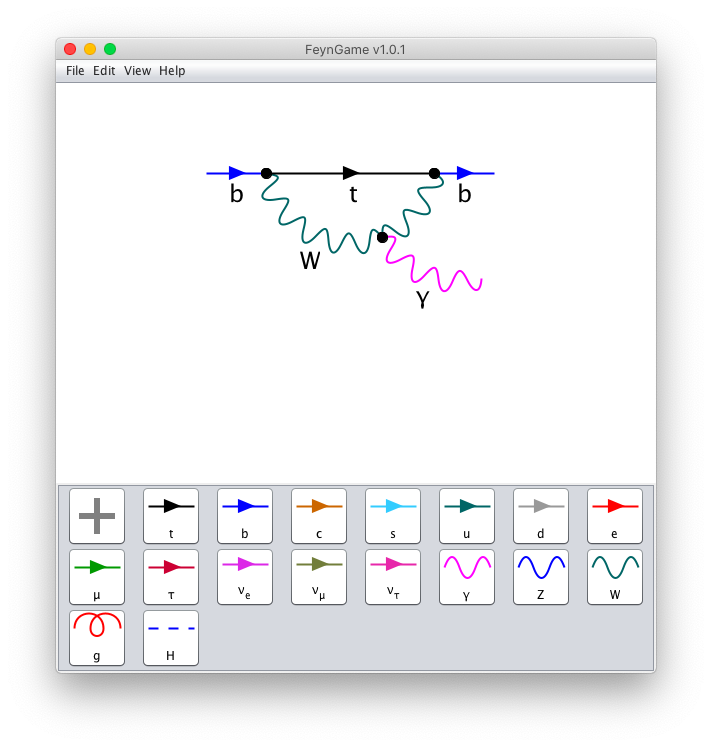} &
           \includegraphics[viewport=60 80 650 580, width=.3\textwidth]{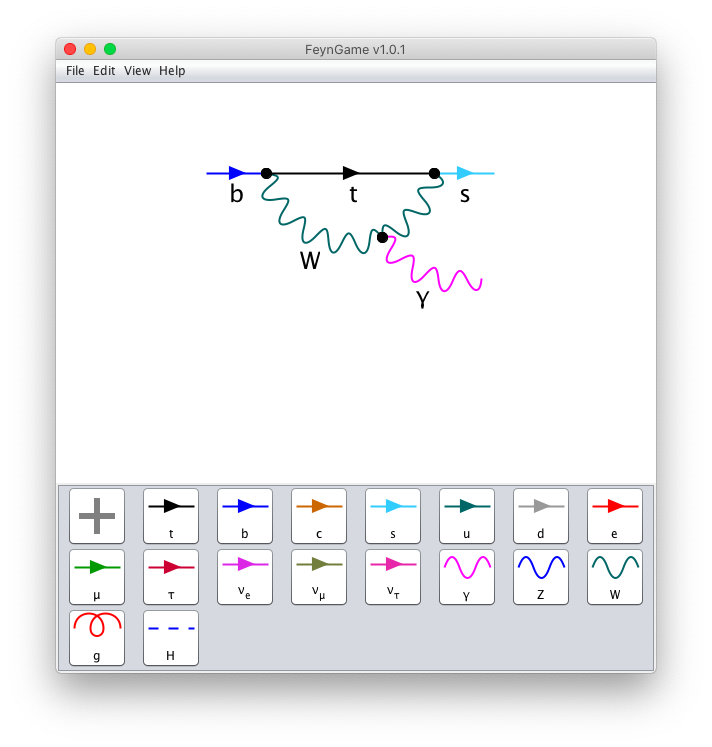} \\
            (iv) & (v) & (vi)
    \end{tabular}
    \parbox{.9\textwidth}{
      \caption[]{\label{fig:b-s-gamma-1}\sloppy  Drawing a diagram for $b\to s\gamma$. Step (i): draw
        a $b$-line. Step (ii): draw a $W$ line. This splits the original
        $b$-line into three. Step (iii): curve $W$ boson line (mouse
        wheel). Step (iv): add a photon line and curve it slightly. Step
        (v): using \editframe, turn the central $b$-line to a
        $t$-line. Step (vi): using \editframe, turn the left $b$-line to
        a $s$ line (The label positions have been slightly adjusted at
        intermediate steps.)  }}
  \end{center}
\end{figure}
%


Another unique feature of \feyngame\ is the automatic splitting of a
line once another line is connected with it through a vertex. This greatly
facilitates the iterative drawing of diagrams, i.e.\ adding lines to
lower order diagrams to create higher order effects through internal
loops or real radiation. Lines can be moved, stretched, or rotated using
a single click-and-drag, the curving of lines is achieved by turning the
mouse wheel, their direction can be changed and labels can be added with
a single keystroke. Almost all line parameters can be adjusted either
through keyboard shortcuts, via a menu (the \editframe), or by editing
the \modelfile.

The state of the canvas (and thus the current diagram) can be stored in
an internal format (\texttt{.fg} files) which includes up to the last
thousand steps in the history of the current state. This allows to
undo/redo any modification of the diagram, no matter when the diagram
was originally created. The current diagram is stored automatically, so
that closing and re-opening \feyngame\ allows one to seamlessly continue
working on the current session.

A diagram can be exported as an image in various formats to a file, it
can be copied to the clipboard (which allows to simply paste it into
presentation tools like \texttt{PowerPoint} or \texttt{Keynote}), and it
can be turned into vector graphics in \texttt{PDF} format, which is
useful for including it in a typeset document as produced by \LaTeX\ or
\texttt{Word}.

The current document gives a first introduction to \feyngame, roughly
following a step-by-step procedure. For simplicity, we will assume that
\feyngame\ was downloaded from
\href{http://www.robert-harlander.de/software/feyngame}{this
  \abbrev{URL}}\footnote{\url{http://www.robert-harlander.de/software/feyngame}}
in the form of a tarball. Unpacking the tarball will produce a folder
named \texttt{feyngame}, with the following structure:
\begin{lstlisting}
   img/          java/         levels/          models/
\end{lstlisting}
We assume that the user's current directory is \texttt{feyngame}.


\section{Draw Mode}\label{eq:draw}


Let us start \feyngame\ without command line arguments and without
loading a \modelfile\ (see \sct{sec:modelfile}).  The call from a
terminal will thus look something like
\begin{lstlisting}
  $ java -jar java/FeynGame.jar
\end{lstlisting}
while on many systems (e.g.\ \texttt{Windows}, \texttt{Linux}) double
clicking \texttt{FeynGame.jar} will work as well.  Either way, this will
open a dialogue window which allows one to choose between ``drawing
mode'' and ``game mode''.  Let us consider drawing mode first.



\subsection{The main window}


%
\begin{figure}
  \begin{center}
    \begin{tabular}{c}
      \includegraphics[viewport=60 80 650 580,
        width=.4\textwidth]{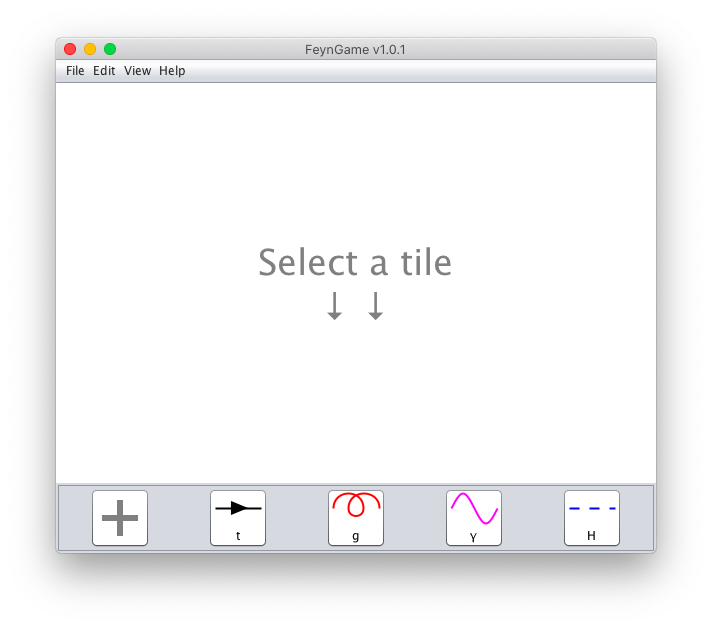}
    \end{tabular}
    \parbox{.9\textwidth}{
      \caption[]{\label{fig:main-window-1}\sloppy  The default main window of \feyngame, showing the
        menu bar with items \menu{File}, \menu{Edit}, \menu{View}, and
        \menu{About}, the canvas below it, and the default model bar at
        the bottom with tiles for the $t$, $g$, $\gamma$, and $H$ line,
        and the ``+'' tile to add new objects to the model.}}
  \end{center}
\end{figure}
%


Selecting the drawing mode, \feyngame\ will open only the ``main
window'', shown in \fig{fig:main-window-1}. It contains the ``canvas''
on which the Feynman diagram will be drawn, and below it a number of
tiles. The latter define graphical objects (the \textit{current model})
which can be used to draw diagrams, except for the left-most tile
(showing a big ``+'') which allows to add new objects to the \textit{current
model} (see \sct{sec:modelfile}). The elements of the menu bar at the top
of the main window will be described in the course of this paper,
whenever the corresponding functionality is discussed.

By default (i.e.\ without loading a \modelfile, see
\sct{sec:modelfile}), \feyngame\ will use a toy model as the
\textit{current model}, consisting of only a few particles which we will
refer to as quark, gluon, photon, and Higgs, respectively (see the tiles
of the main window in \fig{fig:main-window-1}). It is one of the central
ideas of \feyngame\ that this default model can be replaced by a
personalized model via \menu[,]{File, Load model file} in the main menu,
see \sct{sec:modelfile} below. This \modelfile\ will then define the
\textit{current model}.

For the moment, we will assume that the three main \textit{auxiliarly
  features} of \feyngame, described in \rhtab{table:aux} are
de-activated. Thus, if your canvas shows a grid of points, press
\gridkey\ to switch off the \textit{grid} feature for the moment.

\begin{table}
\begin{center}
  \caption{ \label{table:aux}
The three main auxiliary features of \feyngame. They can be
  toggled by pressing the corresponding key, or via the menu item
  \menu{View}, which also allows one to adjust the grid size.
}
\begin{tabular}{|lll|}
  \hline
  feature & key & description\\\hline
  \textit{grid}  & \gridkey& equidistant grid of points on the canvas\\
  \textit{helper lines}&\helperkey
  & indicate the active region of an object\\
  \textit{show active object}
  & \activekey & moving dash pattern for active object on the
  canvas\\\hline
\end{tabular}
\end{center}
\end{table}


\subsection{Single line}\label{sec:single-line}
\begin{figure}
  \begin{center}
    \begin{tabular}{cccc}
      \includegraphics[viewport=60 80 650
        580,width=.22\textwidth]{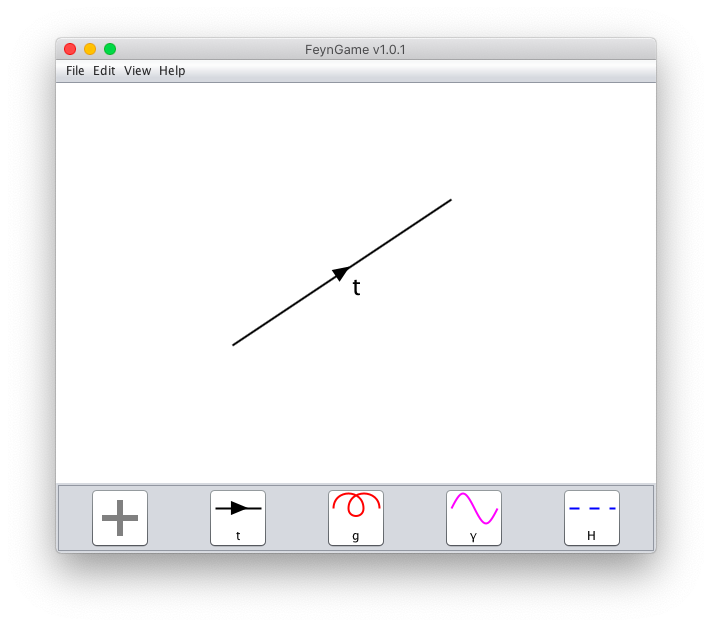} &
      \includegraphics[viewport=60 80 650
        580,width=.22\textwidth]{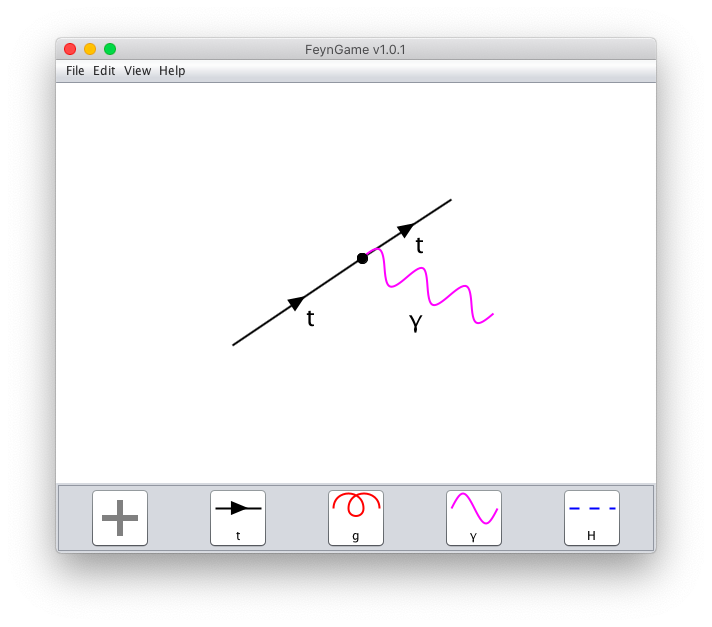}&
      \includegraphics[viewport=60 80 650
        580,width=.22\textwidth]{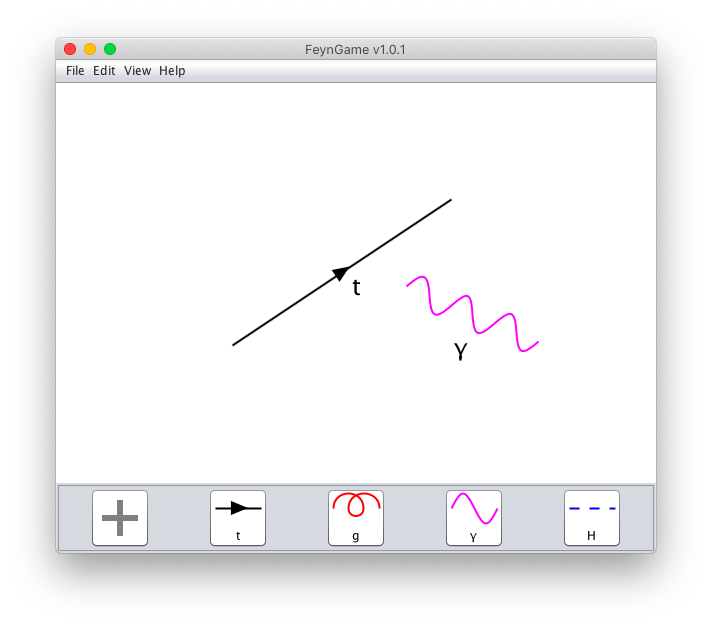}&
      \includegraphics[viewport=60 80 650
        580,width=.22\textwidth]{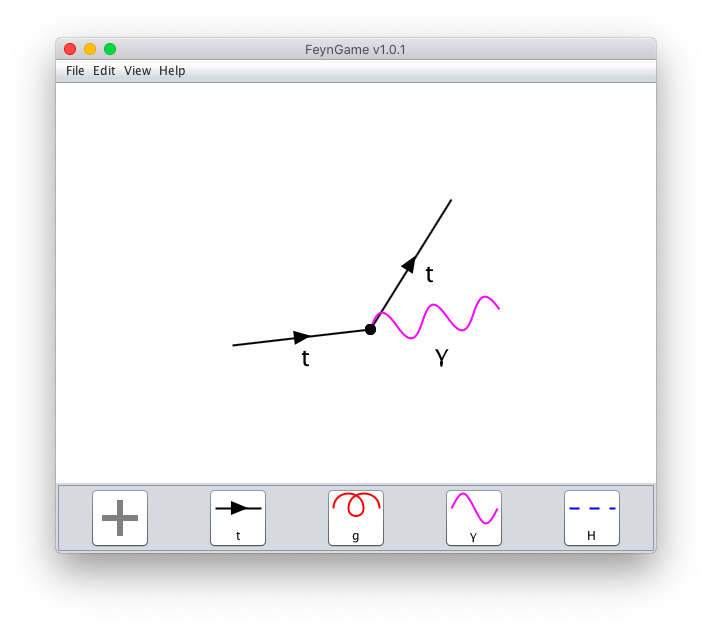}\\
      (a) & (b) & (c) & (d)
    \end{tabular}
    \parbox{.9\textwidth}{
      \caption[]{\label{fig:initial}\sloppy (A) the canvas contains a single $t$ line; (b)
        adding a $\gamma$ line close to it results in two $t$ lines, a
        $\gamma$ line, and a vertex marker; (c) moving the $\gamma$ line
        away from the vertex results in a single $t$ line and a
        single $\gamma$ line; (d) moving the vertex in (b) moves also
        the ends of the lines connected with it. }}
  \end{center}
\end{figure}

Let us select a quark line by a single click on the corresponding
tile. Subsequent click-and-drag on the canvas will draw a straight quark
line between the click and the release point, see
\fig{fig:initial}\,(a). If that line shows a moving dash pattern, press
\activekey\ to switch off the \textit{show active object} feature for
the moment. If you hover the mouse pointer over that line (without
clicking), the width of that line increases slightly; this indicates
that clicking at this point will grab the line. If part of the line
changes color when hovering over it, press \helperkey\ to switch off the
\textit{helper lines} feature for the moment.


%
%

\begin{figure}
  \begin{center}
    \begin{tabular}{c}
      \includegraphics[viewport=0 0 600 500,
        height=.3\textheight]{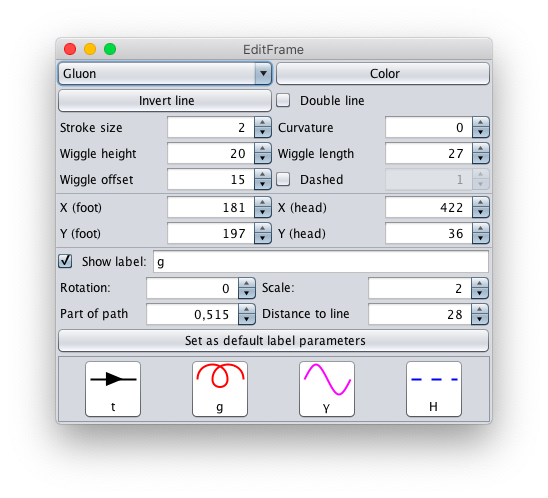}
    \end{tabular}
    \parbox{.9\textwidth}{
      \caption[]{\label{fig:editframe-gluon}\sloppy The \editframe\ window for a gluon line.  }}
  \end{center}
\end{figure}

You can now modify this line in several ways. Let us first discuss
geometrical modifications which are most conveniently applied by
operating directly with the mouse on the canvas:
\begin{description}
\item{\textit{move}:} You can either move the entire line by grabbing
  the line close to its center, or only one end of the line by grabbing
  it in the vicinity of this end. Hovering over the line,
  \feyngame\ will display a small textbox in the upper left corner of
  the canvas which tells you whether grabbing the line at this point
  will move the entire line or just one end of it. Switching on the
  \textit{helper lines} feature (toggle with \helperkey, see
  \rhtab{table:aux}) will indicate this directly by highlighting the
  ``active region'' of the line when hovering over it with the mouse
  pointer.
\item{\textit{curve}:} Turning the mouse wheel will curve the line. Of
  course, after curving, you can again move the line in the same way as
  above. Moving an end of a curved line will preserve the height of the
  corresponding circular segment, which means that the radius of the
  segment will change. This makes it easy to draw closed circles, see
  \sct{sec:circles}.
\item{\textit{invert}:} You can invert the ``direction'' of this line by
  pressing \invertkey. For the quark line, this will invert the
  direction of the arrow on the line. For photons and gluons, it will
  invert the wiggles/spirals. The position of the line label will change
  accordingly (see \sct{sec:labels}).
\end{description}
Other line properties like color, stroke size, or arrow size may
preferably be changed using the \editframe{}. This is a separate window
which can be opened (and closed) by selecting an object on the canvas
and pressing \editframekey, or via the \menu{View} menu item. The actual
content of the \editframe{} window depends on the object that shall be
modified; \fig{fig:editframe-gluon} shows the \editframe{} window for
gluon lines, for example. It also allows to change the relevant
parameters which determine the form of the wiggles. Furthermore, one may
attach a text label which will retain its relative position to the line
even when it is moved or curved, see \sct{sec:labels}.


%
%


For most parameters and options accessible through \editframe{} there
are keyboard shortcuts. A complete list can be displayed through the
menu item \menu[,]{View, Show keyboard shortcuts}. On the other hand,
the position, curvature and orientation of a line can also be controlled
through \editframe{}, which even overrides the presence of the grid.

Any object on the canvas can be removed by selecting it and pressing
backspace, or via the menu item \menu[,]{Edit, Delete active object}.


\subsection{Connecting lines}\label{sec:connect-line}
Let us now pick a photon line by selecting the corresponding tile below
the canvas. Again, click-and-drag will add that line to the canvas. If
one of the end points of the new line is close to the already existing
quark line, the photon line will be connected with it, indicated by a
vertex marker (the black dot), see \fig{fig:initial}\,(b). In this
process, the quark line is split into two adjacent quark lines, each
inheriting the properties of the original quark line (except for the
length, of course). The canvas now contains four objects: two quark
lines, one photon line, and a vertex marker. Clicking on one of these
objects will make it ``active''; the \editframe\ window will always
refer to the active object. Switching on the \textit{show active object}
feature (toggle with \activekey, see \rhtab{table:aux}) will indicate
the active object through a moving dash pattern.

Each of the three lines can be modified in the same way as
above. Removing the photon line from the canvas (see
\sct{sec:single-line}), or detaching it from the vertex will cause the
two quark lines to re-combine to a single quark line if they still have
the same essential properties such as color, width, angle,
etc.\footnote{Text labels will be combined if they differ.} and the
vertex marker will disappear, see \fig{fig:initial}\,(c).

Moving the vertex, on the other hand, will move all ends of the lines
connected with it, see \fig{fig:initial}\,(d). This is a convenient
feature for fine tuning of Feynman diagrams. Other properties of the
vertex that can be modified are accessible by activating it
(i.e.\ clicking on it) and opening \editframe\ (i.e.\ pressing
\editframekey).



\subsection{Model file}\label{sec:modelfile}

\begin{figure}
  \begin{center}
    \begin{tabular}{c}
      \includegraphics[viewport=0 0 600 500, height=.3\textheight]{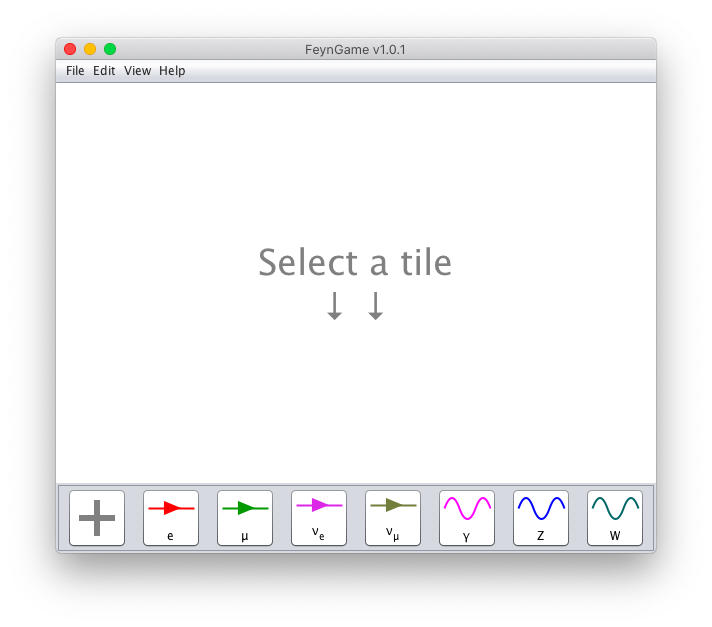}
    \end{tabular}
    \parbox{.9\textwidth}{
      \caption[]{\label{fig:ew2gen}\sloppy \feyngame's main window when called with
        \texttt{EW2gen.model}.  }}
  \end{center}
\end{figure}

As described above, \editframe\ allows one to modify the appearance
of lines and vertices. However, it is more along the philosophy of
\feyngame\ to use a pre-defined model (say, \qed, the \sm, the \mssm, or
a subset of the associated particles) where each particle corresponds to
a line with a certain appearance. For example, if we are only interested
in weak interactions of the first two generations of leptons, the main
window of \feyngame\ could look like \fig{fig:ew2gen}. The leading-order
diagram for muon decay can then be simply drawn by selecting a muon
line, drawing it on the canvas, followed by the line for a $W$ boson, a
muon neutrino $\nu_\mu$, an electron, and an electron neutrino
$\nu_e$. It should not be necessary to change the line styles or
manually add particle labels.


\begin{table}
  \begin{center}
  \caption{\label{table:linestyles} Line types in \feyngame.}
  \begin{tabular}{rc}
    \hline
    \texttt{fermion} & \raisebox{-.3em}{\includegraphics[clip,
      viewport = 18 755 577 830,
    width=.2\textwidth]{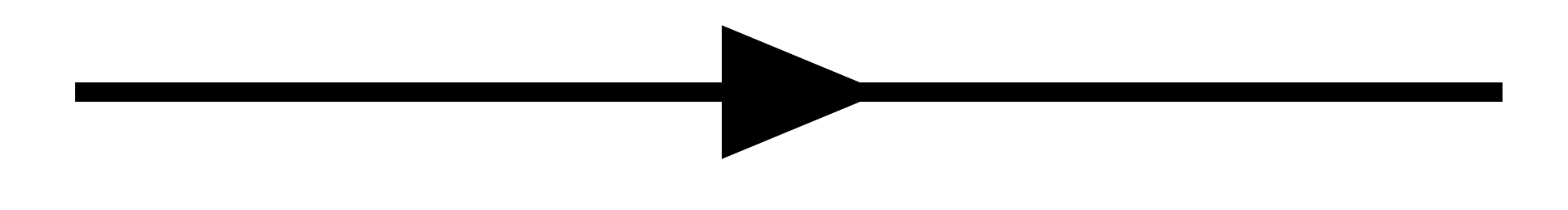}}\\[1em]
  \texttt{photon} & \raisebox{-.5em}{\includegraphics[clip,
      viewport=18 395 363 470,
    width=.2\textwidth]{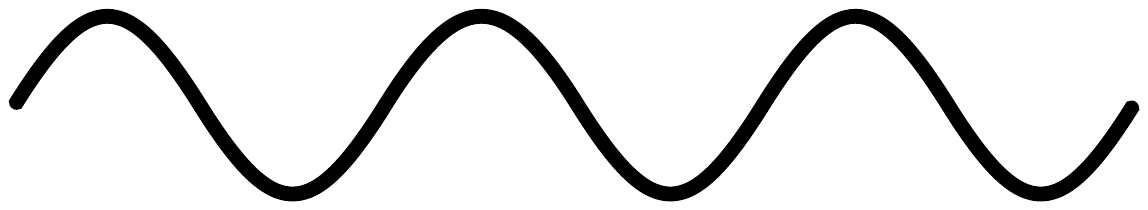}}\\[1em]
  \texttt{gluon} & \raisebox{-.5em}{\includegraphics[clip,
      viewport = 18 405 363 470,
    width=.2\textwidth]{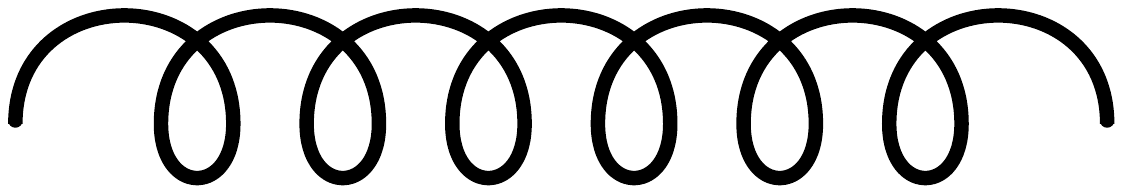}}\\[1em]
  \texttt{scalar} & \raisebox{-.5em}{\includegraphics[clip,
      viewport=18 395 409 470,
      width=.2\textwidth]{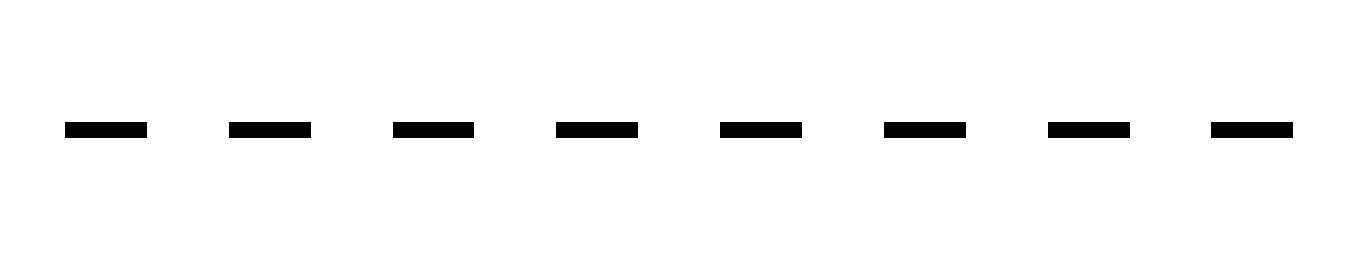}}\\
  \hline
  \end{tabular}
  \end{center}
\end{table}


Drawing a printable version of a diagram thus literally becomes a matter
of seconds, because the different line styles for the individual
particles have been defined beforehand, using the \textit{model
  file}. The \modelfile\ for the example above reads
\begin{lstlisting}
[ e, E, fermion, color=Red, label=e ]
[ nue, Nue, fermion, color=dc27e7, label=<html>&nu<sub>e ]
[ mu, Mu, fermion, color=Green, label=<html>&mu]
[ numu, Numu, fermion, color=27e7b3,label=<html>&nu<sub>&mu ]
[ W, W, photon, color=dc27e7, label=W]
[ Z, Z, photon, color=Blue, label=Z]
[ ph, ph, photon, color=Green, label=<html>&gamma ]
\end{lstlisting}
For future reference, let us assume that these lines are the content of
a file named\break \texttt{EW2gen.model}.  Each pair of square brackets
defines a particle. The first two entries assign an internal name for
the particle. Since fermion lines have a direction, the two entries are
different in this case (e.g.\ \texttt{e,E} for the electron, or
\texttt{nue,Nue} for the electron-neutrino), while they are identical
for bosons (\texttt{W,W} or \texttt{Z,Z}). The third entry defines the
basic line style according to \rhtab{table:linestyles}. These three
entries are required when defining a line in the {\modelfile}. The
\text{label} parameter is optional and will be discussed in more detail
in \sct{sec:labels}.  Other optional parameters are listed in
\rhtab{table:lineoptions}.


\begin{table}
\begin{center}
\caption{\label{table:lineoptions} Line options.}
\begin{tabular}{rll}
  \hline
  option & value & applies to \\\hline
  \texttt{color} & color name, hex & all\\
  \texttt{label} & string, html & all\\
  \texttt{stroke} & pixels & all\\
  \texttt{dash} & true/false & all\\
  \texttt{dashLength} & pixels & all\\
  \texttt{arrowSize} & pixels & \texttt{fermion}\\
  \texttt{wiggleSize} & pixels & \texttt{photon, gluon}\\
  \texttt{wiggleHeight} & pixels & \texttt{photon, gluon}\\
  \texttt{wiggleOffset} & pixels & \texttt{gluon}\\
  \texttt{wiggleSharpness} & pixels & \texttt{photon}\\
  \texttt{double} & true/false & all\\\hline
\end{tabular}
\end{center}
\end{table}


%
%


Starting \feyngame\ by giving it the name of the \modelfile\ as argument
(including its path) will load the specific model instead of the default
one. For example, if we say
\begin{lstlisting}
  $ java -jar java/FeynGame.jar models/EW2gen.model
\end{lstlisting}
\feyngame\ will directly start in drawing mode, using the model defined
in \texttt{EW2gen.model} as the \textit{current model}, with the main
window as shown in \fig{fig:ew2gen}. Alternatively, one can change the
\textit{current model} from the main menu of \feyngame\ using
\menu[,]{File, Load model file}.

Aside from the default, built-in model, \feyngame\ comes with model
files for \qed, the \sm, and also with \texttt{EW2gen.model}, see
\fig{fig:ew2gen}. The user may adjust the parameters of these models to
ones personal taste simply by editing the corresponding model
file. Another, sometimes more convenient option is to modify the
appearance of an existing line within \feyngame\ (for example by using
\editframe), and then selecting the ``+'' tile in the main window. This
will add a new object tile to the \textit{current model} at the bottom
of the main window. The new line is thus available as a new object in
\feyngame. Selecting \menu[,]{File, Save model file (as)} will modify
the current \modelfile\ (or create a new one), so that the new line is
available also in future sessions. Similarly, one may remove existing
object tiles by \keystroke{\shift}-clicking the tile and
subsequently saving the \modelfile.

A third way to modify the \modelfile, which is a mixture of the two
options just discussed, will be described in \sct{sec:copypaste}.


\subsection{Vertices}\label{sec:vertices}

\begin{figure}
  \begin{center}
    \begin{tabular}{cc}
      \includegraphics[viewport=60 80 650
        580,width=.4\textwidth]{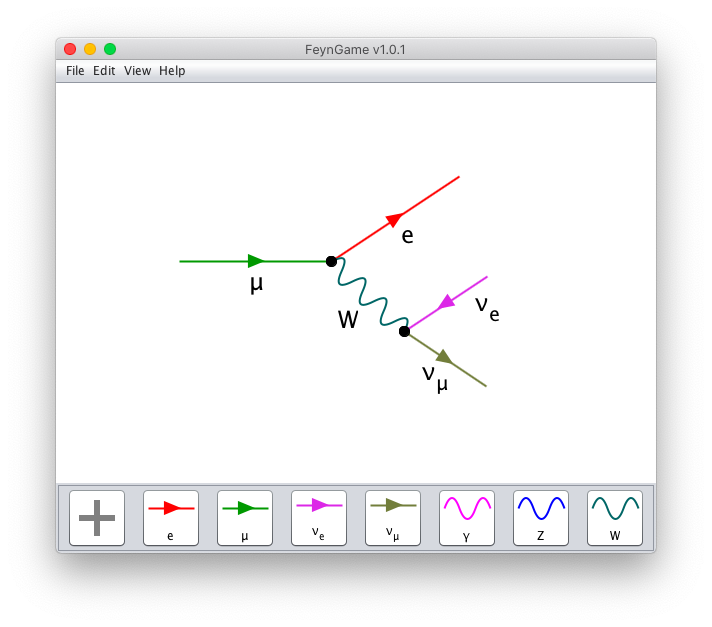} &
      \includegraphics[viewport=60 80 460
        220,width=.4\textwidth]{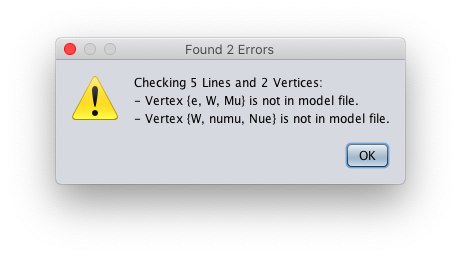}\\
      (a) & (b)
    \end{tabular}
    \parbox{.9\textwidth}{
      \caption[]{\label{fig:diacheck}\sloppy (a) A ``wrong'' diagram for $\mu$ decay. (b)
        Pressing \keystroke{f} reports the errors.  }}
  \end{center}
\end{figure}

Optionally, you may also specify the vertices of a model in the model
file. Unless you assign specific markers to these vertices (see
\sct{sec:vertexmarkers} below), this has no effect on the actual drawing
or appearance of a diagram. However, if the vertices of the model are
specified, you can ask \feyngame\ to check whether the specific diagram
on the canvas is consistent with the current model. For example, assume
that you add the following lines to \texttt{EW2gen.model}:
\begin{lstlisting}
{e, E, ph}
{mu, Mu, ph}
{e, E, Z}
{mu, Mu, Z}
{nue, Nue, Z}
{mue, Mue, Z}
{e, Nue, W}
{nue, E, W}
{mu, Numu, W}
{numu, Mu, W}
{W, W, ph}
{W, W, Z}
\end{lstlisting}
This tells \feyngame\ the topological Feynman rules for the electro-weak
interaction of the first two lepton generations. It will still allow you
to draw any Feynman diagram you like, for example the one shown in
\fig{fig:diacheck}\,(a). However, by pressing \keystroke{f},
\feyngame\ will report that this diagram is not consistent with the
current model by displaying the message in \fig{fig:diacheck}\,(b).


%
%


The main purpose of defining the vertices of a model is in the context
of the game mode of \feyngame\ though, see \sct{sec:game}. However,
aside from the simple consistency check described here, it may also be
useful to assign specific vertex markers to some of the vertices. This
is described in the next section.


\subsection{Vertex markers}\label{sec:vertexmarkers}

\begin{figure}
  \begin{center}
    \begin{tabular}{c}
      \includegraphics[viewport=60 80 650
        580,width=.5\textwidth]{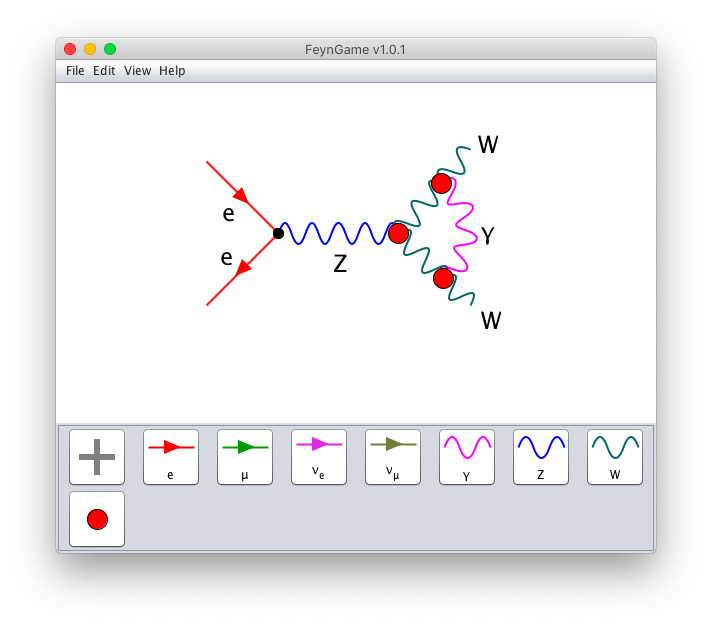}
    \end{tabular}
    \parbox{.9\textwidth}{
      \caption[]{\label{fig:vmarker}\sloppy  Triple gauge boson vertices are marked in red.  }}
  \end{center}
\end{figure}

As described in \sct{sec:connect-line}, \feyngame\ draws a vertex marker
at the point where lines are connected with one another. Using
\editframe, the style of this marker can be modified. By selecting the
``+'' tile, an object tile for the currently active vertex marker will
be created. Similar to newly created lines, the new marker style can
even be saved to the \modelfile\ using \menu[,]{File, Save model file},
so that it is available as a \textit{free-floating object} for future
sessions when calling \feyngame\ with this file (see below). The
\modelfile\ format for a marker is
\begin{lstlisting}
(vname=redmark, size=10, borderColor=ff000000, borderDashed=false, borderDashLength=5, borderStroke=1, fillingColor=Red)
\end{lstlisting}
This would define a big red vertex marker in \feyngame.  The meaning of
the parameters should be self-explanatory. If the parameter
\texttt{vname} is missing or is equal to the string
``\texttt{default}'', \feyngame\ will use this marker as
default.\footnote{If there are multiple vertices without the
  \texttt{vname} parameter or it being equal to ``default'', the
  uppermost of these vertex definitions is used as the default vertex.}
Otherwise, it will use the internal default marker. On the other hand,
you can assign a specific marker to an individual vertex by adding the
attribute \texttt{type} to the vertex definition. For example, modifying
the triple gauge boson vertices defined above in \texttt{EW2gen.model}
according to
\begin{lstlisting}
{W, W, ph, type=redmark}
{W, W, Z, type=redmark}
\end{lstlisting}
will use the red marker for these vertices, and the default marker for
all others, see \fig{fig:vmarker}.


%
%


The new marker will now also appear as a tile object in the
\feyngame\ main window. It can be placed anywhere on the canvas as a
\textit{free-floating object}, in the sense that it will not be connected
with any other object (it does ``clip'' to grid points though, see
\sct{sec:grid}).

\feyngame\ provides a set of more elaborate vertex markers. They are
accessible through the drop-down menu \menu{Select filling from default
  patterns} of the vertex marker \editframe, or by adding the
\texttt{fxpath} option to the vertex definition in the \modelfile. For
example,
\begin{lstlisting}
(fxpath = cross.fx)
\end{lstlisting}
will provide a vertex marker with a cross in the middle. A list of
available vertex markers and their corresponding \texttt{fxpath}
parameter are shown in \rhtab{table:fx}.

\begin{table}
\begin{center}
\caption{\label{table:fx} JavaFX vertex markers.}
\begin{tabular}{rc}
  \hline
  \texttt{fxpath} & vertex marker\\\hline
  \texttt{cross.fx} & \raisebox{-.2em}{\includegraphics[viewport = 18 265 577 874,width=1em]{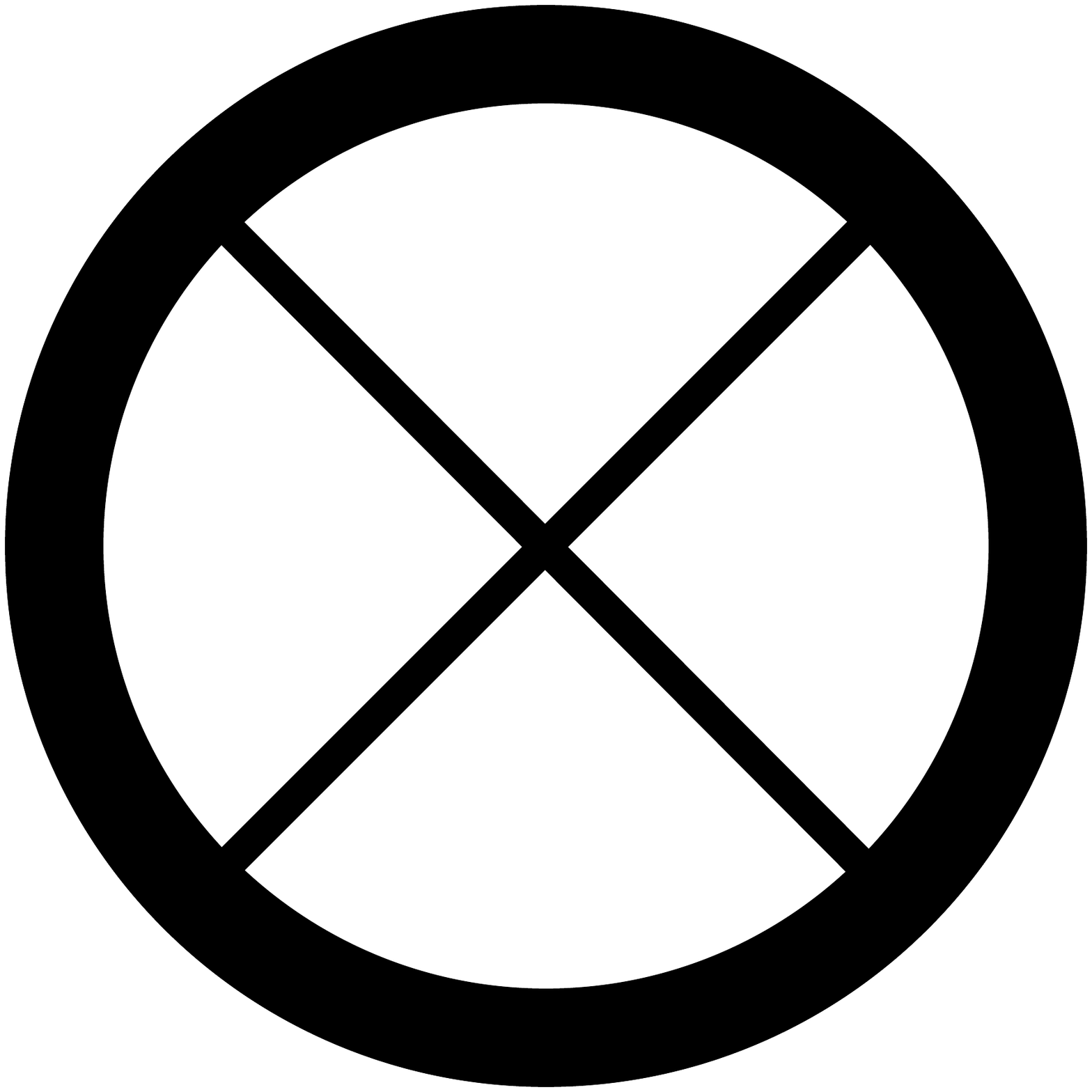}}\\
  \texttt{hatched.fx} & \raisebox{-.2em}{\includegraphics[viewport = 18 265 577 874,width=1em]{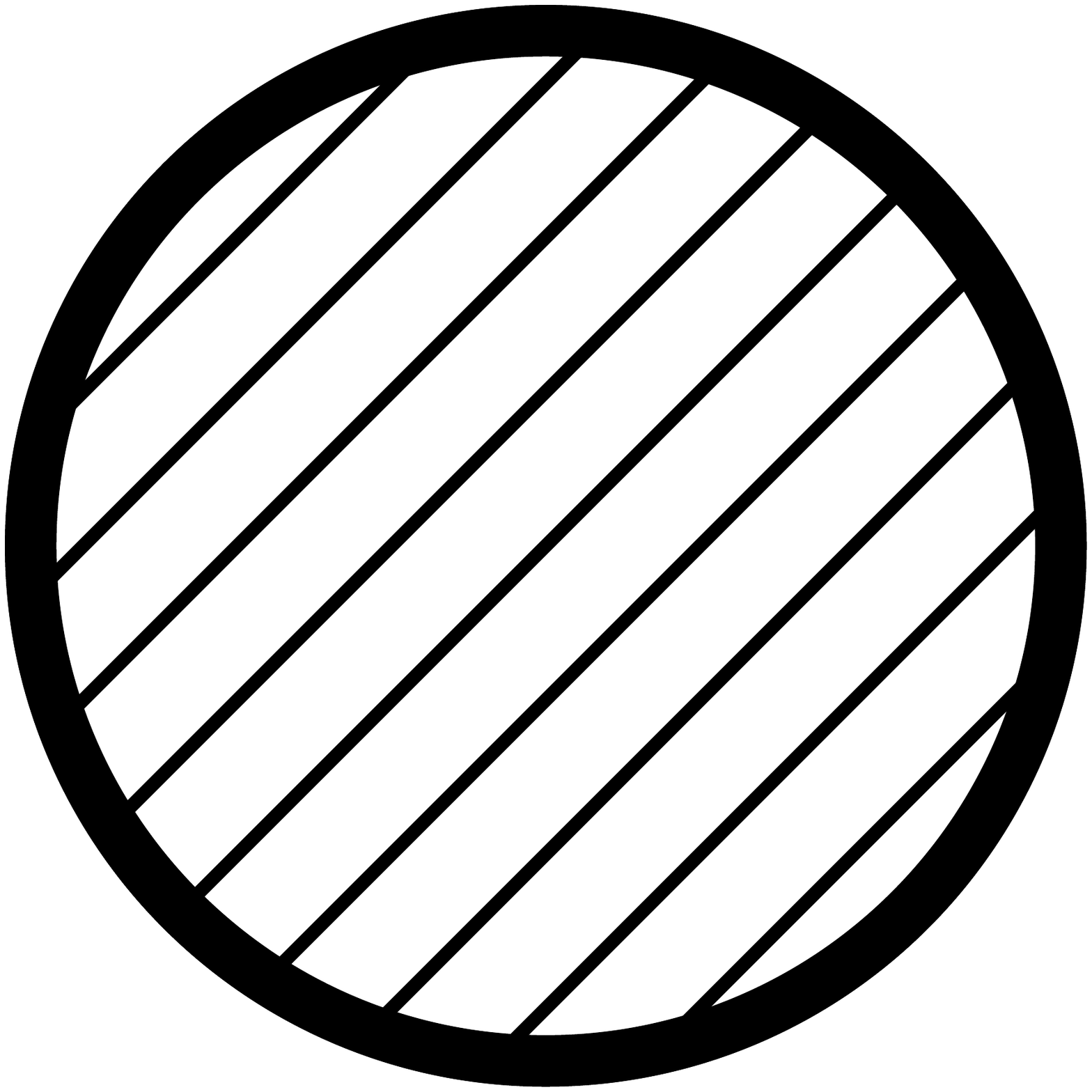}}\\
  \texttt{crosshatched.fx} & \raisebox{-.2em}{\includegraphics[viewport = 18 265 577
    874,width=1em]{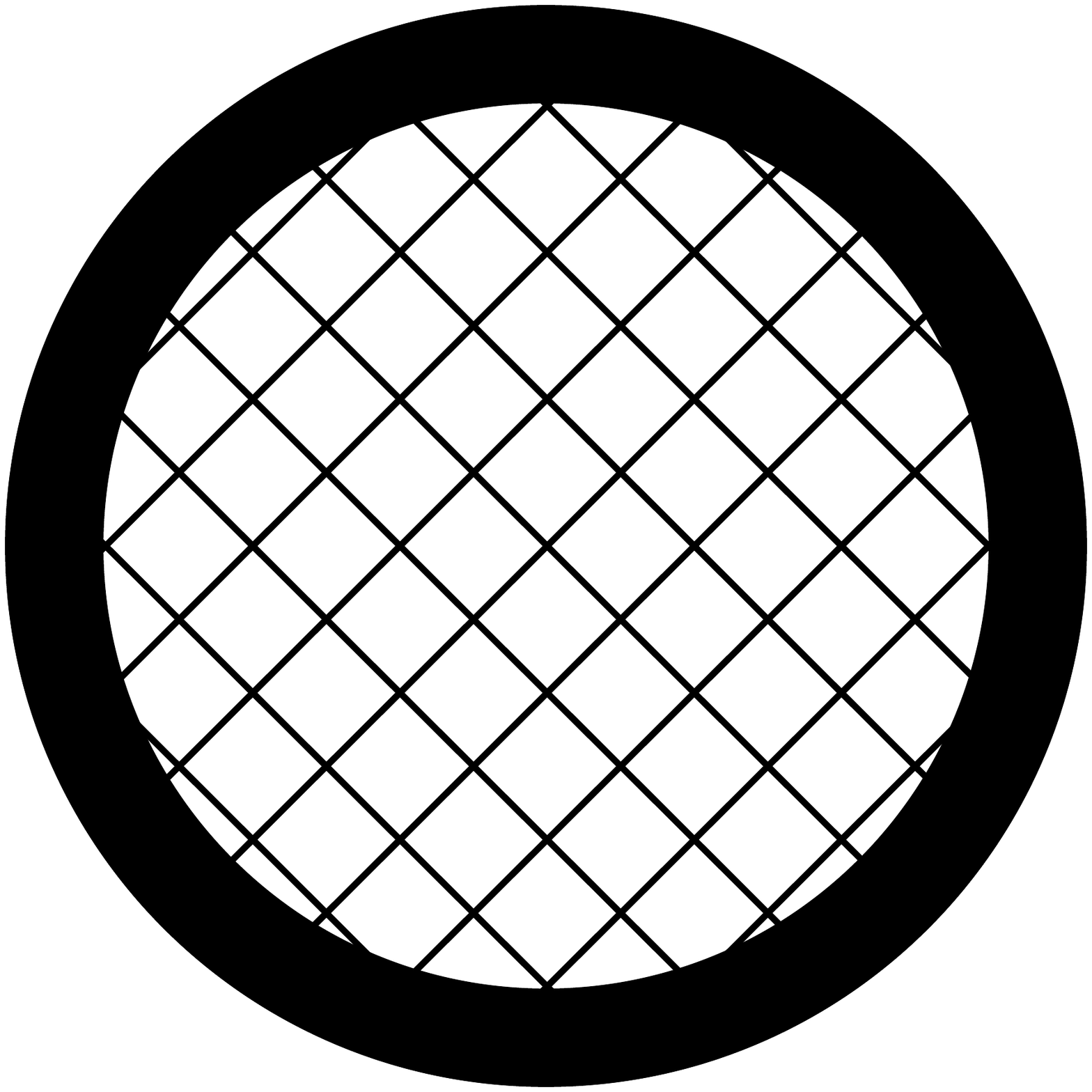}}\\
  \texttt{star.fx} & \raisebox{-.2em}{\includegraphics[viewport = 18 265 577
    874,width=1em]{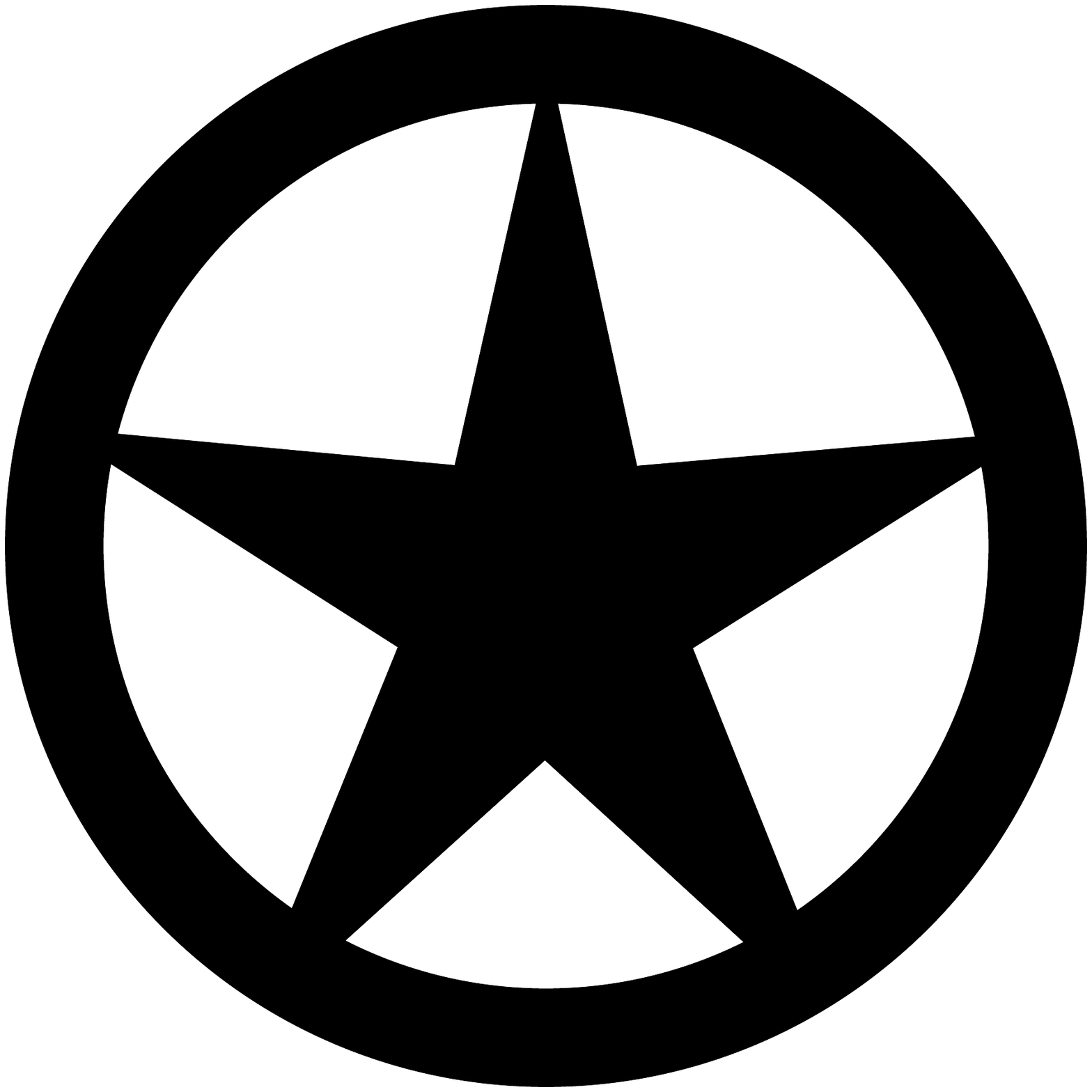}}\\
  \hline
\end{tabular}
\end{center}
\end{table}


\subsection{Labels}\label{sec:labels}

As indicated in \sct{sec:modelfile}, it is possible to attach labels to
lines. In the \modelfile, this is done by including the
\texttt{label=<label>} keyword in the line definition, see
\texttt{EW2gen.model}. Currently, labels can be given in plain text
format, or using basic \abbrev{HTML} commands. In this case, the label
definition should start with the string ``\texttt{<html>}'' (see the
definition of the $\nu_e$ label in \texttt{EW2gen.model}, for
example). Using the \texttt{<html>} combined with the \texttt{<font>} tag,
it is possible to change the color and the font of the text.

The position of a label is relative to its associated line, but it can
be changed by mouse or using \editframe. Moving the line will also move
the label, such that the relative position of the two remains fixed. On
the other hand, the orientation of the label is defined relative to the
canvas, i.e.\ rotating the line will not rotate the label relative to
the canvas.  The \textit{default} properties of the label (position,
size, orientation, etc.) are the same for all lines; they can be changed
by pressing the \menu{Set as default} button in \editframe. In this way,
one can also determine whether the label is shown or not by
default. Pressing \keystroke{l} will toggle the displaying of the active
line's label.

In the same way, labels can also be attached to vertex markers and other objects.


\subsection{Other features}


\subsubsection{Grid}\label{sec:grid}

If the grid is switched on (toggle with \gridkey, see
\rhtab{table:aux}), all objects except for line labels will be
``clipping'' to this grid.\footnote{This applies only to newly
  positioned objects. The position of object already on the canvas will
  not be altered by switching on the grid.} This means that the
endpoints of lines, or the center of \textit{free-floating objects} (see
\sct{sec:vertexmarkers}), can only be placed at the grid points.  The
grid points are shown on the canvas, but are not included when the
diagram is exported as an image or a \texttt{PDF}. This also holds for
the other auxiliary options like \textit{helper lines} or \textit{show
  active object}.  The spacing of the grid points can be changed from
the menu \menu[,]{View, In(De)crease grid spacing}, or through the
associated keyboard shortcut, see \menu[,]{View, Show keyboard
  shortcuts}.

In addition to the visible grid points on the canvas, every line on the
canvas introduces a ``local'' set of grid points
along this line, roughly at the same distance as the visible grid points on the
canvas. If the \textit{grid} is active, lines can be connected with one
another only at these line-specific grid points.

Sometimes one may want to have an object to be located off the
grid. This can be achieved either by temporarily switching off the grid
(e.g.\ by pressing \gridkey). An object placed without the grid will
remain at its (possibly off-grid) position even after the grid is
switched back on. The other option to place an object off the grid is
via \editframe, which allows one to specify the coordinates of the
object in pixels.


\subsubsection{Clipping}\label{sec:clipping}

Independent of whether the grid is on or off, \feyngame\ will do
additional kinds of clipping when objects are moved by mouse or curved
by mouse wheel:
\begin{description}
\item[Vertex clipping.] If the end of a line is moved closer to another
  line (or to its other end) than a certain minimal distance, it will be
  connected with that line through a vertex.
\item[Curvature clipping.] If the curvature of a line is decreased below
  a certain minimal value via the mouse wheel, it will be set to zero.
\item[Angle clipping.] If the slope of a straight line is close to a
  multiple of $\pi/2$, it will be replaced by that multiple of
  $\pi/2$. Via the menu item \menu[,]{Edit, Clipping angles} or by
  repeatedly pressing \keystroke{c}, that value can be changed to
  $\pi/4$ or $\pi/16$, or zero.
\end{description}
This kind of clipping is introduced for the user's convenience, because
otherwise it would be rather clumsy to connect lines, or to turn a
curved line into a completely straight line. In order to overrule such
clipping, one can use \editframe\ to do a pixel-wise modification of the
position or the curvature.


\subsubsection{Copy/Paste}\label{sec:copypaste}

Objects can be copied and pasted using \menu[,]{Edit, Copy} and
\menu[,]{Edit, Paste} from the main menu, or through the system-defined
keyboard shortcut (\keystroke{\cmd-c}/\keystroke{\cmd-v} for
\texttt{MacOS}, for example). The object will be pasted to the center of
the canvas (if pasted from menu) or close to the mouse pointer (if
pasted from keyboard).

One may also copy an object from the canvas to the clipboard, and then
paste it to a regular text editor. It will then display the model-file
format of that object. Vice versa, one may paste any line from the model
file into the \feyngame\ canvas, and it will show the graphical
representation of that object.


\subsubsection{Text}

There is no specific text object in \feyngame. As described above,
labels of lines and vertices are attributes of these objects. If one
really needs a separate text element which is not associated with any
visible object, one may define a vertex of size zero, and introduce the
text as the vertex label, e.g.
\begin{lstlisting}
(vname=text, size = 0, label=mytext)
\end{lstlisting}


\subsubsection{Images}

There are different ways to include an image on the canvas. The first
one is to define it as a vertex marker, for example by opening
\editframe\ for an arbitrary vertex on the canvas, and clicking
\menu[,]{Select image / JavaFX file}, which opens the file
chooser, from where one may select an arbitrary image (\texttt{JPG},
\texttt{PNG}, \texttt{GIF} or \texttt{BMP}). One can then include it in the
\textit{current  model} with the ``+'' tile. Or one defines it directly
in the model file; the syntax is:
\begin{lstlisting}
(imagePath=<path_to_image>/Unicorn.jpg)
\end{lstlisting}
Other options like \texttt{size}, \texttt{rotation}, etc.\ are available
but not shown here. Note that this method puts the image into a round
shape, which means that part of it may be cut away. However, one can use
this object like an other vertex marker; for example, one may associate
it with a specific vertex as described in \sct{sec:vertexmarkers}.

Another way to include an image is to simply paste it from the clipboard
into the \feyngame\ canvas, using the system-defined keyboard
shortcut. For example, in \texttt{MacOS} one could use
\keystroke{\cmd-\ctrl-\shift-4} to take a screen shot and paste it into
the \feyngame\ canvas using \keystroke{\cmd-v}. Again, the image can be
added as an object to the \textit{current model} using the ``+''
tile. When stored to the model file, it will appear there as a line of
the following form:
\begin{lstlisting}
|imagePath=<path>/<image>.png, rotation=0, scale=1|
\end{lstlisting}
The options should be self-explanatory.


\subsubsection{Circles and Tadpoles}\label{sec:circles}


%
\begin{figure}
  \begin{center}
    \begin{tabular}{cccc}
      \includegraphics[viewport=60 80 650
        580,width=.22\textwidth]{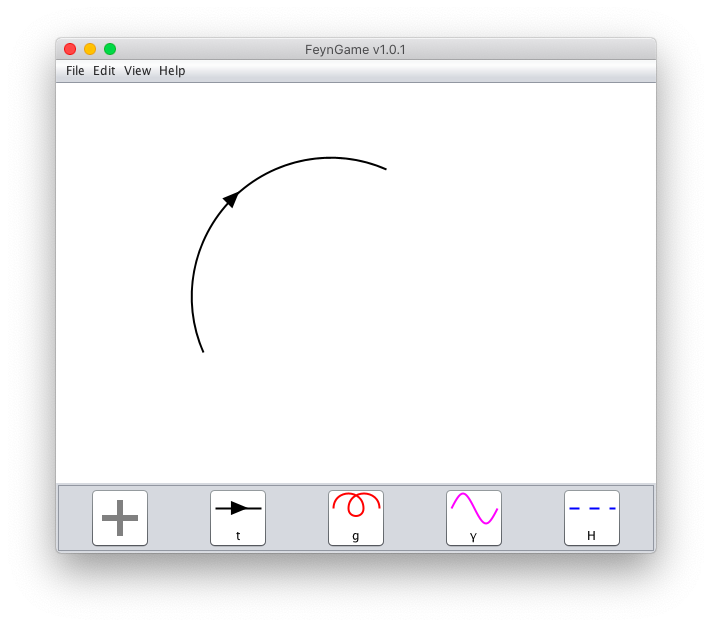} &
      \includegraphics[viewport=60 80 650
        580,width=.22\textwidth]{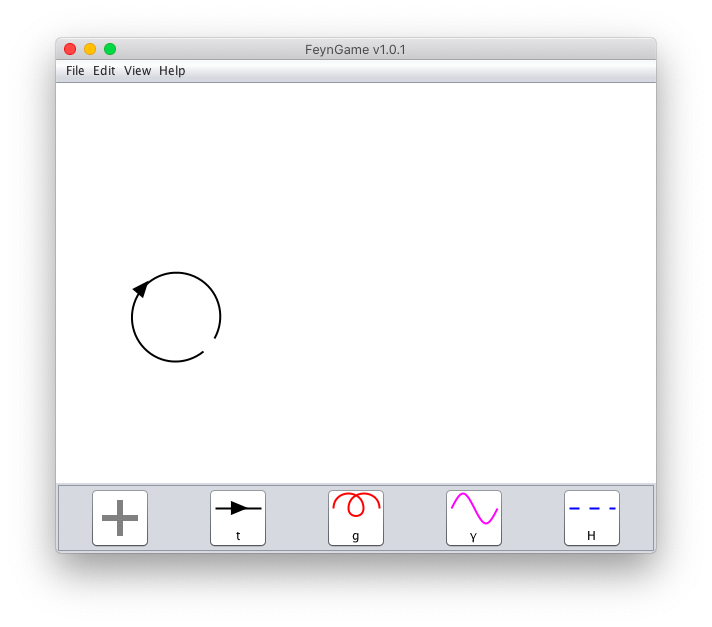} &
      \includegraphics[viewport=60 80 650
        580,width=.22\textwidth]{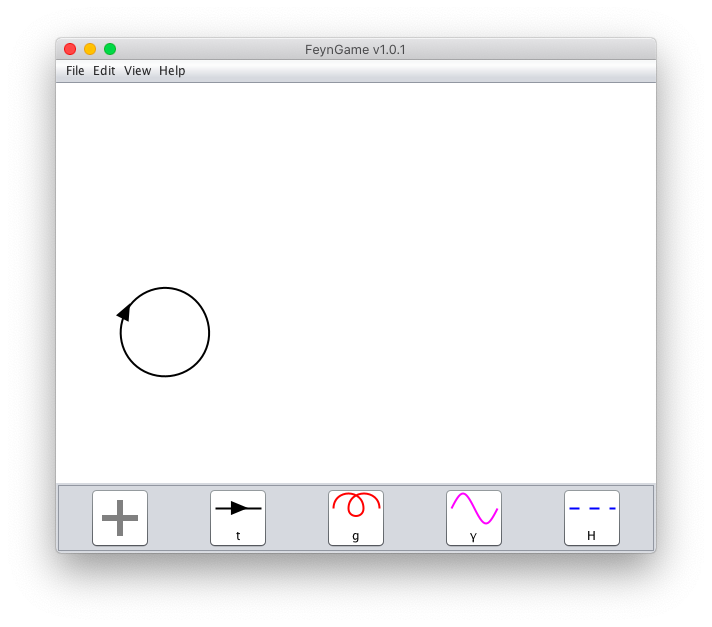} &
      \includegraphics[viewport=60 80 650
        580,width=.22\textwidth]{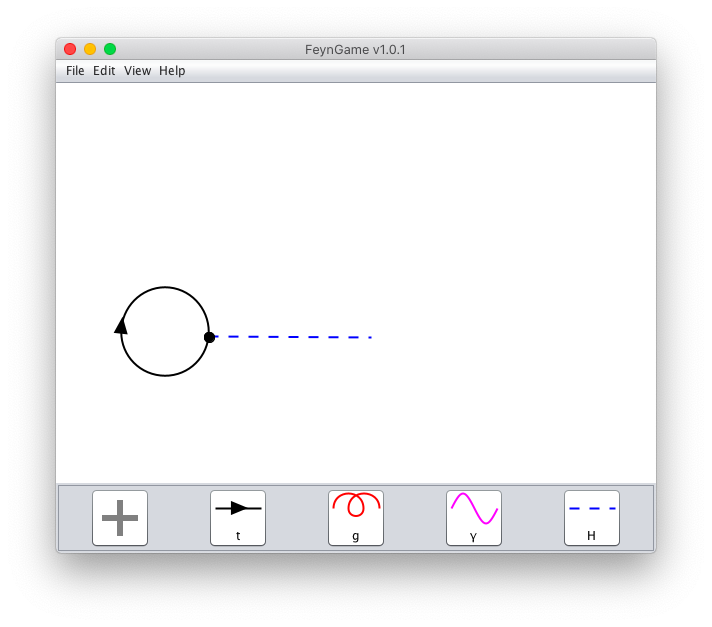}
    \end{tabular}
    \parbox{.9\textwidth}{
      \caption[]{\label{fig:tadpole}\sloppy A circle can be drawn by identifying the ends of a
        curved line (a)--(c). Attaching a line to the circle turns it
        into a tadpole (d); the vertex becomes the new ``glue point''. }}
  \end{center}
\end{figure}
%


We define a circle diagram as a closed loop which is not connected with
any other line. It can be obtained by connecting the ends of a curved
line with each other, see \fig{fig:tadpole} (a)--(c). One can ``re-open''
the circle and turn it back to a regular curved line simply by
click-and-drag in the vicinity of the ``glue point''. Once we connect a
(single) line with the circle, it becomes a tadpole diagram, see
\fig{fig:tadpole}\,(d). While the actual position of the original circle
does not change, its glue point gets identified with the new vertex.



\subsection{Saving and Exporting}\label{sec:saveexport}

\paragraph{Saving a diagram.} \feyngame\ automatically saves the status
of the canvas across sessions in an internal format (\texttt{.fg} file),
including its history.\footnote{Up to the last thousand steps.} This
means that if you open \feyngame, the session will start at the point
where it was previously closed.  Selecting \menu[,]{File, New} from the
menu, on the other hand, will clear the canvas and the history.

Selecting \menu[,]{File, Save (as)} from the menu allows you to save the
status of the canvas, including its history, in the internal format to a
specific file. You can modify the diagram at a later stage by opening it
in the canvas using \menu[,]{File, Open}. Selecting \menu[,]{Edit,
  Delete history} will delete the history of the canvas. Storing a
diagram without history typically reduces the size of the
\texttt{.fg}-file significantly.

\paragraph{Export as image.} A diagram can be exported to an image file
in various formats using \menu[,]{File, Export as image} from the
menu. None of the auxiliary features from Table\,\ref{table:aux} will be
exported, only the plain diagram will be visible. Exporting as
``Portable Network Graphics ($\ast$.png)'', the background of the
diagram will be transparent. In addition, the diagram can be exported to
the clipboard using \menu[,]{File, Export to clipboard}. In this case,
the background of the image will be transparent. This is particularly
useful for including the diagram into a \texttt{PowerPoint} or
\texttt{Keynote} presentation.


%
\begin{figure}
  \begin{center}
    \begin{tabular}{cc}
      \raisebox{0em}{%
        \includegraphics[%
          height=.6\textwidth]%
                        {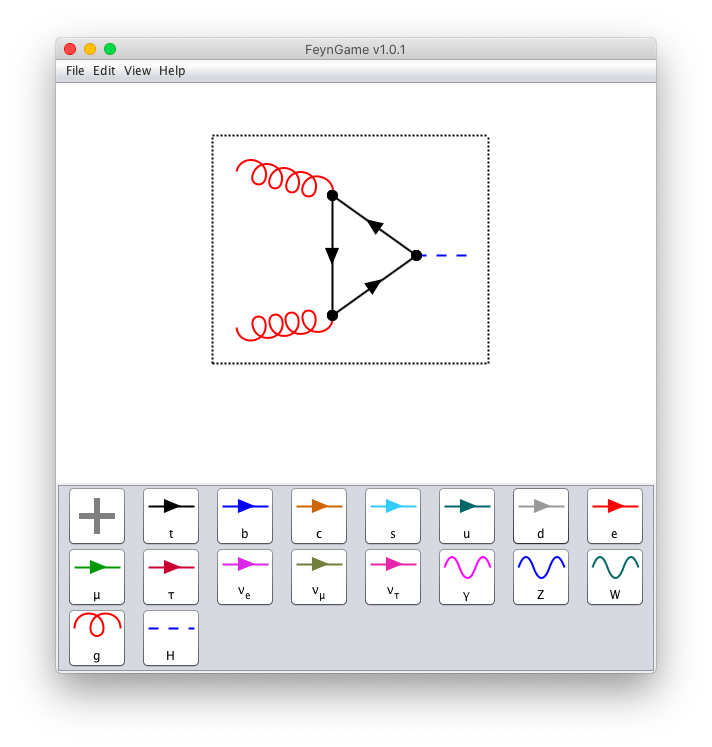}} &
      \raisebox{0em}{%
        \includegraphics[%
          height=.4\textwidth]%
                        {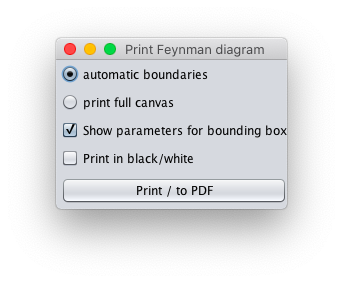}} \\[-2em]
      (a) & (b) \\
      \multicolumn{2}{c}{\includegraphics[%
          height=.3\textwidth]%
        {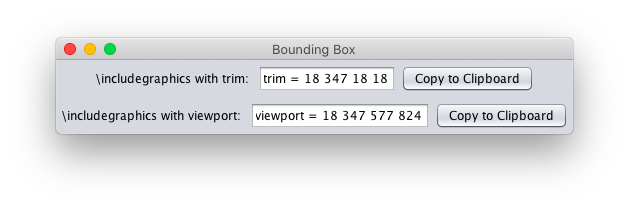}}\\[-3em]
      \multicolumn{2}{c}{(d)}
    \end{tabular}
    \parbox{.9\textwidth}{
      \caption[]{\label{fig:print}\sloppy
        Printing to \pdf. (a) Diagram to be printed using
        \menu[,]{File,Print / to PDF}; the bounding box is indicated as
        a black dotted frame in this picture. (b) Dialogue box with
        bounding box and black-and-white options. (c) Bounding box information.
    }}
  \end{center}
\end{figure}
%


\paragraph{Exporting as PDF/printing.} With \menu[,]{File, Print / to PDF}, a
diagram can be exported as \pdf, or directly sent to a
printer. \feyngame\ will open a dialog box which controls the following
features, see \fig{fig:print}\,(b):
\begin{itemize}
\item \textit{automatic boundaries:} The bounding box of the \pdf\ is
  determined automatically from the actual location of the diagram on
  the canvas, roughly as indicated by the dotted black frame around the
  diagram in \fig{fig:print}\,(a).
\item \textit{print full canvas:} The bounding box is determined by the
  size of the canvas.
\item \textit{Show parameters of bounding box:} The bounding box
  parameters are displayed when the diagram is printed or exported, see
  \fig{fig:print}\,(c). They can then be copied to the clipboard and
  pasted into a text editor, for example as parameters for the
  \verb$\includegraphics$ command in \LaTeX:
  \begin{lstlisting}
  \includegraphics[viewport = 18 347 577 824,
    size=.3\textwidth]{ggh.pdf}\end{lstlisting}%
  Equivalently, one can use the \texttt{trim} parameters, see
  \fig{fig:print}\,(b).
\item \textit{Print in B/W:} Print in black-and-white. More precisely:
  Lines and the borders of vertices are drawn in black.  The interior of
  the vertices is drawn black or white depending on the brightness of
  its actual color. All other objects remain untouched.
\end{itemize}


\subsection{Misc}
\begin{description}
\item[Object layering:] Lines on the canvas are layered according to
  their activation history, i.e.\ the line that was active last is at
  the top, etc. This also holds for their labels. Vertex markers,
  including their labels, are always on top of everything.
\item[Canvas shifting:] \keystroke{\shift}-Click-and-drag on empty space on the canvas will
  move the whole diagram.
\item[Right-clicking:] Right-clicking anywhere on the canvas opens a
  menu next to the mouse pointer which provides an alternative to
  selecting objects by clicking on the tiles in the main
  window. Right-clicking close to an object in addition provides the
  possibility to change the style of this object, as an alternative to
  using the \editframe.
\end{description}



\section{Game mode}\label{sec:game}


\subsection{Idea of the game}


%
\begin{figure}
  \begin{center}
    \begin{tabular}{cc}
      \includegraphics[viewport=60 80 650 780,
        width=.4\textwidth]{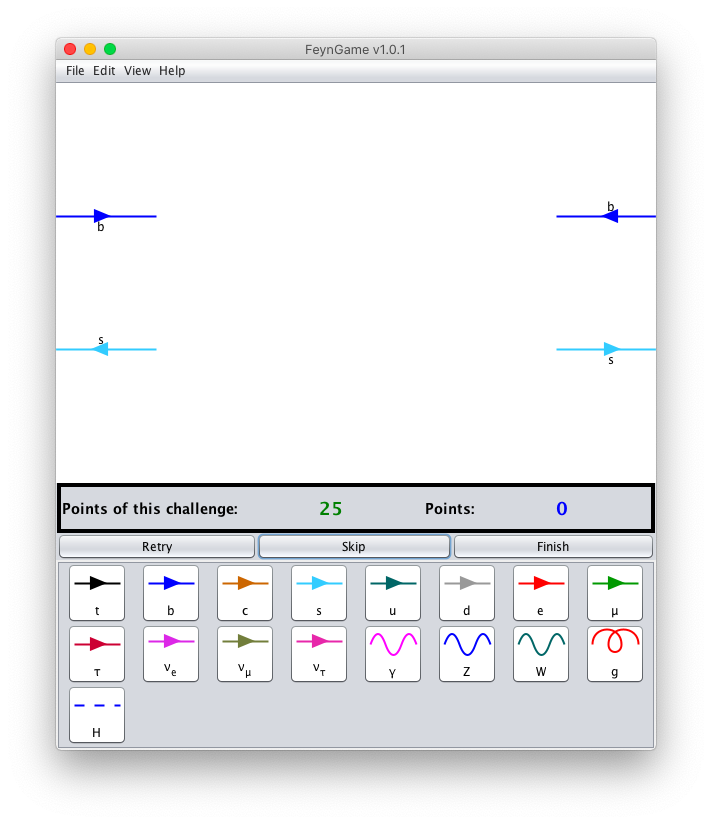} &
      \includegraphics[viewport=60 80 650 780,
        width=.4\textwidth]{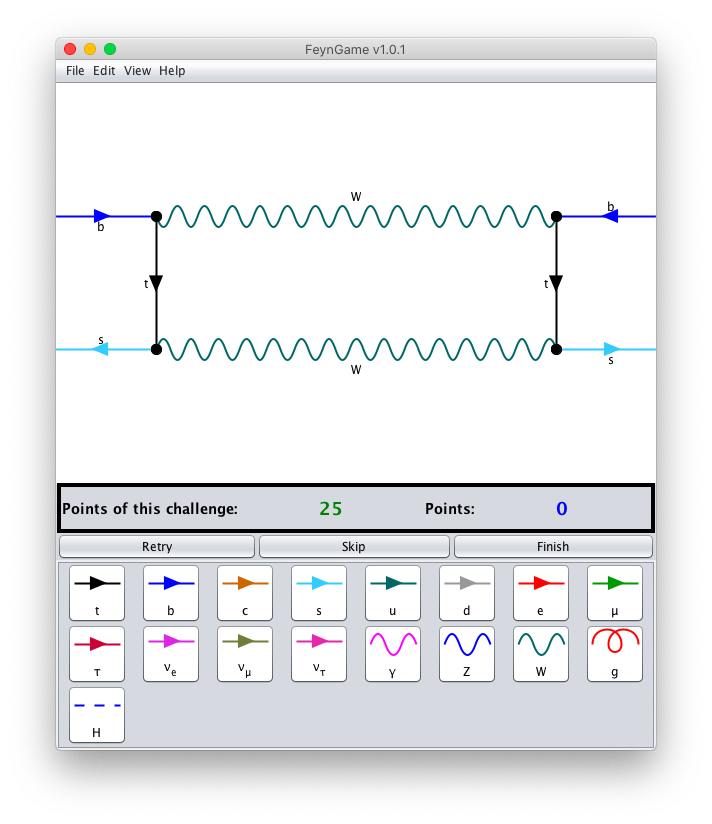}\\ (a) & (b)
    \end{tabular}
    \parbox{.9\textwidth}{
      \caption[]{\label{fig:game}\sloppy \feyngame\ in game mode. (a) the ``challenge'' is to
        connect the initial (left) with the final state (right) through a
        connected Feynman diagram which only uses lines and vertices of
        the underlying model. (b) a possible solution.}}
  \end{center}
\end{figure}
%

Currently, \feyngame\ provides a single game mode, called \infin\ (for
\underline{In}itial-\underline{Fin}al). It can be started by passing a
\levelfile\ to \feyngame\ upon start-up:
\begin{lstlisting}
java -jar java/FeynGame.jar levels/SMLevel.if
\end{lstlisting}
Alternatively, one can start \feyngame\ without any argument, choose
\texttt{InFin} from the dialogue window, and subsequently specify the
\levelfile. With the \levelfile\ \texttt{SMLevel.if} provided with the
\feyngame\ distribution, the main window typically looks similar to
\fig{fig:game}. The lines on the canvas indicate the initial and the
final state of a process. The goal is to draw a connected Feynman
diagram that contributes to this process, using the Feynman rules of the
current model (the \sm\ in this case). An example for a valid diagram is
shown in \fig{fig:game}\,(b). Pressing \menu{Finish} will add the
``Points of this challenge'' as indicated below the canvas to the total
number of points. If the diagram is not valid, pressing \menu{Finish}
will report the errors.  Choosing ``Use a timer with duration in
seconds'' in the dialogue box upon starting \feyngame\ in game mode will
restrict the total available time for the game to the specified number
of seconds. This can also be achieved by passing the number of seconds
as an integer to \feyngame\ when calling it from the command line like
above.

When drawing the diagram, it is important to note that the outer end of
each external particle is tied to the border of the window and cannot be
moved; the other end has to be connected with some other particle
(external or internal). Also, this is \textit{the only way} to connect
an external particle with the diagram, i.e., other particles cannot be
connected with external lines anywhere else but at this ``inner end''.


\subsection{Level file}

A sample \levelfile\ looks like this:
\begin{lstlisting}
model = ../models/standard.model

easy
start: e,E
end: e,E
start: e,e
end: e,e

medium
start: H
end: Z,g,g
start: mu
end: e,numu,Nue

hard
start: b
end: s,ph
start: b,S
end: B,s
\end{lstlisting}
Of course, \feyngame\ needs to know the underlying model. For that
purpose, one needs to point to a proper \modelfile, including the vertex
information, whose structure is as discussed in \sct{sec:modelfile} and
\ref{sec:vertices}. This is done in line~1; the path to the file can be
relative to the path of the \levelfile\ or absolute.\footnote{As usual,
  the home directory can be abbreviated by the character
  ``{\raise.17ex\hbox{$\scriptstyle\sim$}}'' on Linux-like systems.} The
rest of the \levelfile\ is divided into three blocks, headed by the
keywords ``\texttt{easy}'', ``\texttt{medium}'', and
``\texttt{hard}''. Below each of these headings is a list of processes,
specified by their initial state and final state (\texttt{start} and
\texttt{end}, respectively). Easy/medium/hard processes are assigned
5/15/25 points.

The user may modify the \levelfile\ by adding or removing processes,
changing the model file, etc. In principle, given a \modelfile, the
generation of the \levelfile\ could be automated, of course; this is
currently work in progress.


\subsubsection{Directories and Files}

As indicated above, \feyngame\ automatically stores the current state of
the canvas anytime during the session. It will report the location of
the corresponding files upon exit. For example, in \texttt{MacOS}, it
prints the following message on the terminal:
\begin{lstlisting}
Saved preferences to: <home>/Library/Preferences/FeynGame/DrawMode.ini
Saved diagram to: <home>/Library/Preferences/FeynGame/last.fg
\end{lstlisting}
While \texttt{last.fg} is a binary file which contains the current
status of the canvas, including its history, \texttt{DrawMode.ini} is a
regular \texttt{ASCII} file which lists the current parameters of the
session. They will be applied upon the next start of \feyngame. While it
is not recommended, one may thus modify the parameters in
\texttt{DrawMode.ini} using a text editor. However, it is usually
simpler (and safer) to adjust the global parameters from within \feyngame.


\section{Conclusions and Outlook}\label{sec:outlook}

The purpose of \feyngame\ is two-fold: On the one hand, it should be a
useful tool for the efficient and intuitive drawing of Feynman diagrams
for presentations and publications. On the other hand, it serves a
didactic purpose, in the sense that it should playfully convey the
concept of Feynman diagrams to non-experts. The basis for this is its
ability to check the validity of a certain Feynman diagram w.r.t.\ an
underlying model. The current version of \feyngame\ contains one such
game, called \texttt{InFin}, which is close to one of the main actual
applications of Feynman diagrams, namely the theoretical description of
a scattering or decay process. An example for another possible game is a
Scrabble-type game where valid Feynman diagrams must be created from
randomly emerging lines.

Other future plans for \feyngame\ include the im- and exporting of
Feynman diagrams to text-based programs like
\texttt{TikZ-Feynman}\,\cite{Ellis:2016jkw}, and the possibility to
include \LaTeX\ text.

\feyngame\ is published as open source under the GNU General Public
License. It can be downloaded from
\url{https://gitlab.com/feyngame/FeynGame}, or as a gzipped tar ball
from \url{https://web.physik.rwth-aachen.de/~harlander/software/feyngame}.


\paragraph{Acknowledgments.}
We would like to thank Jonas Klappert, Fabian Lange, Magnus Schaaf, Nils
Sch\"oneberg, and Ma\l gorzata Worek for helpful comments.


\newcommand{\bibentry}[4]{#1, {\it #2}, #3\ifthenelse{\equal{#4}{}}{}{, }#4.}
\newcommand{\journal}[5]{\href{https://dx.doi.org/#5}{\textit{#1} {\bf #2} (#3) #4}}
\newcommand{\arxiv}[2]{\href{https://arXiv.org/abs/#1}{\texttt{arXiv:#1\,[#2]}}}
\newcommand{\arxivhepph}[1]{\href{https://arXiv.org/abs/hep-ph/#1}{\texttt{hep-ph/#1}}}
\newcommand{\arxivhepth}[1]{\href{https://arXiv.org/abs/hep-th/#1}{\texttt{hep-th/#1}}}
\newcommand{\arxivheplat}[1]{\href{https://arXiv.org/abs/hep-lat/#1}{\texttt{hep-lat/#1}}}
\newcommand{\arxivmathph}[1]{\href{https://arXiv.org/abs/math-ph/#1}{\texttt{math-ph/#1}}}
\newcommand{\arxivmath}[1]{\href{https://arXiv.org/abs/math/#1}{\texttt{math/#1}}}
\IfFileExists{./\jobname_ref.tex}{%

}{}


\end{document}